\documentclass[twocolumn,aps]{revtex4-2}
\usepackage{graphicx}
\usepackage{amssymb}
\usepackage{amsfonts}
\usepackage{amsmath}
\usepackage{amsbsy}
\usepackage{float}
\usepackage{natbib,mhchem}
\usepackage{comment}
\usepackage{soul}
\usepackage{color}
\usepackage[dvipsnames]{xcolor}
\usepackage[colorlinks=true, urlcolor=blue, citecolor=blue, linkcolor=blue]{hyperref}

\newcommand{\be}{\begin{equation}}
	\newcommand{\ee}{\end{equation}}
\newcommand{\ben}{\begin{eqnarray}}
	\newcommand{\een}{\end{eqnarray}}
\begin{document}
%
	
	\title{
Opto- and magneto-tunable exceptional degeneracies in non-Hermitian ferromagnet/$p$-wave magnet junctions}

	\author{Mohammad Alipourzadeh}

\affiliation{
	Department of Physics, Faculty of Science, Shahid Chamran University of Ahvaz, 6135743135 Ahvaz, Iran 
	}

			\author{Davood Afshar}

		\affiliation{
			Department of Physics, Faculty of Science, Shahid Chamran University of Ahvaz, 6135743135 Ahvaz, Iran 
			}
			
			\author{Yaser Hajati}

		\affiliation{
			Department of Physics, Faculty of Science, Shahid Chamran University of Ahvaz, 6135743135 Ahvaz, Iran 
			}

\date{\today}
\begin{abstract} 
Unconventional $p$-wave magnets (UPMs) with odd-parity spin textures have attracted interest for their zero net magnetization and anisotropic spin-split Fermi surfaces. Here, we explore a non-Hermitian open quantum system composed of a ferromagnet and a UPM, subjected to an external magnetic field and off-resonant circularly polarized light (CPL), serving as tunable control parameters. We demonstrate the emergence of exceptional points (EPs) in the proposed junction, whose locations can be modulated by the intrinsic properties of the UPM. These EPs exhibit different multiplicities and formation conditions compared to those in even-parity magnets (dubbed $d$- wave altermagnets), a distinction attributable to the preserved time-reversal and broken inversion symmetries characteristic of UPMs. We find that both the unidirectional magnetic field (with adjustable strength and orientation) and the CPL induce momentum-direction-dependent modifications to the EPs, such as their shifting, tilting, merging, or annihilation, supported by analyses of spin projection and eigenvector overlap. Although both perturbations influence the EP structure, they operate via distinct mechanisms: CPL induces a global Floquet re-normalization, enabling dynamic tunability through light, whereas the unidirectional magnetic field selectively alters orientation-aligned terms, lacking such tunability. Beyond revealing EP dynamics in UPM-based junctions, our results highlight UPMs as promising platforms for non-Hermitian phenomena in future spintronics.
\end{abstract}

\maketitle
\section{Introduction}

Magnetic materials are of interest for both fundamental characteristics and technological applications\cite{vzutic2004spintronics,eschrig2015spin,baltz2018antiferromagnetic}. While ferromagnets (FMs) with spin polarization and non-zero magnetization and antiferromagnets (AFMs) with antiparallel magnetic moments have been regarded the main magnetic phases, recent studies have revealed a distinct phase, where spin-split bands emerge despite compensated magnetic order \cite{vsmejkal2022emerging, hellenes2023exchange}. This phase exemplified by materials like RuO$_2$ and MnTe, which commonly termed altermagnets (AMs) \cite{liao2024separation,krempasky2024altermagnetic,lee2024broken, vsmejkal2022beyond, vsmejkal2022giant}.

Magnetic materials can classified based on their spin group symmetries \cite{vsmejkal2022emerging}. In AMs with even-parity symmetry, encompassing $d$-, $g$-, and $i$-wave classifications \cite{ezawa2025third}, inversion symmetry dictates that opposite spins on symmetrically related sublattices are coupled, while time-reversal symmetry (TRS) is broken. This broken TRS leads to the lifting of spin degeneracy and the emergence of anisotropic band structures in reciprocal space \cite{krempasky2024altermagnetic}, mathematically expressed as $E_\sigma(\textbf{k}) = E_{\sigma}(-\textbf{k})$ and $E_\sigma(\textbf{k}) \neq E_{-\sigma}(\textbf{k})$, where $\sigma$ denotes the spin state and $\textbf{k}$ is the wave vector \cite{hedayati2025transverse,krempasky2024altermagnetic}

On the other hand, magnets with odd-parity symmetry, commonly called unconventional $p$-wave magnets (UPMs) \cite{fukaya2024fate}, exemplified by the CeNiAsO compound \cite{hellenes2023exchange}, exhibit preserved TRS but broken inversion symmetry. This results in the inverse relationship $E_\sigma(\textbf{k}) \neq E_{\sigma}(-\textbf{k})$ and $E_\sigma(\textbf{k}) = E_{-\sigma}(-\textbf{k})$ \cite{hedayati2025transverse,hellenes2023exchange}. The significant anisotropy observed in the band structures of UPMs has attracted considerable attention due to their potential for applications in diverse electronic and spintronic systems \cite{fukaya2025josephson2,hedayati2025transverse,maeda2024theory,soori2025crossed}. $p$-wave magnets can host collinear and noncollinear odd-parity spin textures, however, here we focus on the collinear case, where spin- and phase-dependent hopping on a square lattice breaks inversion symmetry.\cite{hedayati2025transverse}.

 Hermiticity forms the foundation of the Hamiltonian description for isolated quantum systems. However, real-world systems are typically open, experiencing some degree of coupling to the environment, which introduces dissipation and complicates their theoretical treatment. While approaches such as the Lindblad quantum master equation offer a rigorous framework for such systems \cite{lindblad1976generators}, its mathematical intricacy limits its applicability. As a more tractable alternative, non-Hermitian (NH) effective Hamiltonians have proven valuable in diverse settings, including dissipative optics, mechanics, and open quantum systems \cite{el2018non,lu2014topological}.
There has been growing theoretical and experimental interest in NH systems in recent years \cite{ashida2020non,kawabata2019symmetry,bergholtz2021exceptional,banerjee2023non}, owing to their unconventional features—such as the NH skin effect \cite{li2023enhancement,zhang2022review}, nontrivial topological behavior \cite{bergholtz2021exceptional}, and extensions of Bloch band theory \cite{yokomizo2019non}, which are fundamentally inaccessible in Hermitian frameworks.

A salient feature of NH models is the emergence of complex energy spectra exhibiting degeneracies termed exceptional points (EPs) \cite{heiss2012physics,dembowski2001experimental,lee2009observation,arouca2023topological}. In contrast to Hermitian systems, NH systems allow for the coalescence of both eigenvalues and eigenstates at EPs. While initially considered mathematical curiosities, EPs are now recognized as topologically significant entities, enabling exotic phases that lack counterparts in Hermitian physics \cite{ashida2020non,cayao2023exceptional}. EPs have already played a pivotal role in the discovery of unconventional topological effects \cite{ashida2020non}, including ultra-sensitive detection techniques \cite{hodaei2017enhanced,chen2017exceptional}, unidirectional lasing \cite{longhi2017unidirectional}, and bulk Fermi arcs \cite{delplace2021symmetry,nagai2020dmft,zhou2018observation,yoshida2018non,okugawa2019topological}, all of which are unrealizable within Hermitian systems.

The intriguing characteristics of EPs have stimulated extensive research across various physical systems, encompassing AMs \cite{reja2024emergence,dash2024role}, topological insulators \cite{bergholtz2019non}, semiconductors \cite{cayao2023exceptional}, semimetals \cite{dey2024hall}, and superconductors \cite{cayao2024non,cayao2022exceptional,kornich2022andreev,kornich2023current,kornich2022signature}. For instance, Cayao demonstrated the emergence of highly tunable EPs along momentum-space rings in a FM/semiconductor junction in the presence of Rashba spin-orbit coupling (RSOC) \cite{cayao2023exceptional}. Furthermore, Cayao and Black-Schaffer explored an open FM-superconductor system, revealing that non-Hermiticity alone induces odd-frequency pairing in these systems \cite{cayao2022exceptional}. Very recently, emerging of N\'eel vector controlled exceptional contours as rings and lines has reported in UPMs \cite{reja2025n}.

While EPs have been investigated in even-parity AM open systems \cite{reja2024emergence,dash2024role}, their tunability in the odd-parity UPMs remains largely unexplored. To address this gap, our work demonstrates EPs in an NH FM/UPM junction, revealing how their positions are governed by both intrinsic UPM properties and external control parameters. Unlike $d$-wave AM systems, these EPs exhibit unique multiplicities and conditions stemming from TRS coexisting with broken inversion symmetry. We establish that a three-dimensional magnetic field (through its strength and orientation) and off-resonance circularly polarized light (CPL) via Floquet engineering of virtual photon processes, synergistically enable unprecedented control over EP dynamics, including momentum-direction-dependent shifting, tilting, merging, and annihilation. Spin projection textures and eigenvector overlap analyses confirm these observations and also show that planar configurations of these controls produce emergent superposition effects beyond their individual capabilities.
The synergy of these controls unlocks transformative functionalities in NH spintronics: (1) dynamically reconfigurable EP-based sensors with enhanced sensitivity, and (2) Floquet-induced topological phase transitions. These advances not only deepen our understanding of odd-parity magnetic systems but also establish them as versatile platforms for active light-controlled quantum materials with precisely engineerable EPs, paving the way for next-generation quantum applications.

\section{Theoretical model}

\begin{figure}
\includegraphics[scale=0.74]{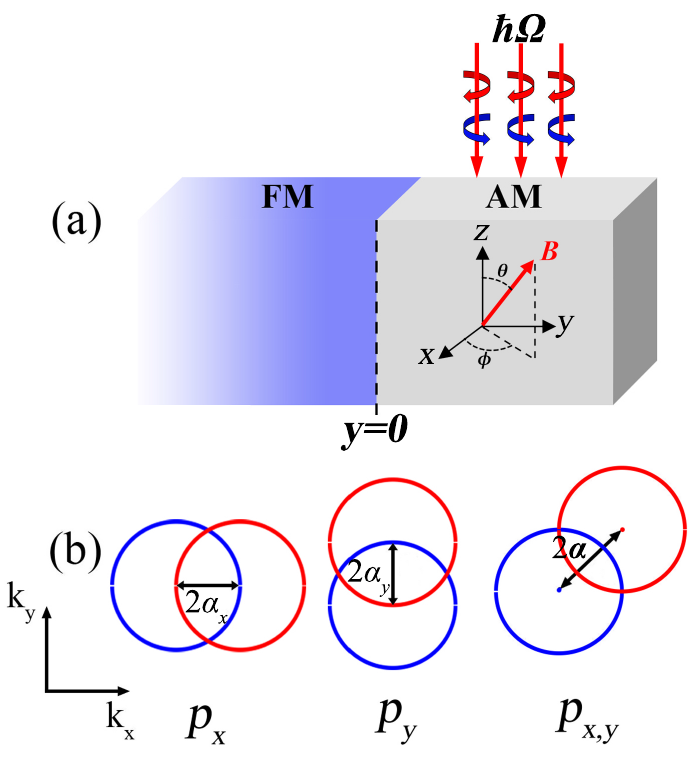}
\caption{Schematic illustration of (a) proposed NH FM/UPM open system in the presence of magnetic field and CPL, and (b) the Fermi surfaces of unperturbed UPM in different sub-waves. The blue and red arrows in (a) illustrates the right- or left-handed circular polarizations and, $\theta$ and $\phi$ refer to the polar and azimuthal angles of the magnetic filed vector, respectively. Different colors of Fermi circles in (b) indicate different spin states.   } 
\label{Fig1}
\end{figure}

Prior to a detailed exposition of the model and its resulting phenomena, it is pertinent to analyze the schematic representation of the UPM-based proposed junction. Figure \ref{Fig1}(a) illustrates the proposed structure, depicting a UPM coupled to a semi-infinite FM lead. The interface between the two regions is located at $y=0$, with the FM lead occupying the region $y<0$ and the UPM located at $y>0$. An external magnetic field ($\boldsymbol{B}$) is also introduced, with the assumption that its orientation can be along the $x$-, $y$-, or $z$-direction, denoted as $B_x$, $B_y$, and $B_z$, respectively. The magnetic field can be defined on the basis of the polar ($\theta$) and azimuthal ($\phi$) angles as $\textbf{B}=B (\sin (\theta) \cos (\phi), \sin(\theta) \sin (\phi), \cos(\theta))$, such that $\theta \in (0,\pi)$ and $\phi \in (0,2\pi)$. In the rest, $B_{x,y,z}$ denote the energy associated with the magnetic fields and we have chosen $\mu_B=1$ for simplicity. Also, the off-resonance CPL with tunable right- or left-handed circularization is applied to the system.

The unperturbed form of the proposed open NH system depicted in Fig. \ref{Fig1}(a) can be effectively modeled by the following Hamiltonian \cite{reja2024emergence,cayao2023exceptional}
\begin{equation}
H=H_p+H_R+\Sigma^r,
\label{Eq1}
\end{equation}
where $H_p$ is the Hermitian Hamiltonian describing the closed UPM system, given by \cite{hedayati2025transverse,maeda2024theory}
\begin{equation}
H_p = t \left( (\mathbf{k}^2 + \boldsymbol{\alpha}^2) \sigma_0 + 2 \mathbf{k} \cdot \boldsymbol{\alpha} \, \sigma_z \right) - \mu \sigma_0,
\label{Eq2}
\end{equation}
in which $\mathbf{k}=(k_x, k_y)$ is the wave vector with the magnitude of $\sqrt{k_x^2+k_y^2}$, $\boldsymbol{\alpha}=(\alpha_x,\alpha_y)$ is the magnetization vector of the UPM with the magnitude of $\sqrt{\alpha_x^2+\alpha_y^2}$, $\mu$ represents the chemical potential, and $t=\hbar^2/2m$ is the hopping parameter, that fixed at 1 during this work. The terms $\sigma_0$ and $\sigma_{x,y,z}$ denote the identity and spin Pauli matrices, respectively.

The term $H_R$ in Eq. (\ref{Eq1}) represents the contribution from RSOC arising due to the proximity of the lead, which is considered FM layer in this work and is expressed as \cite{cayao2022exceptional,reja2024emergence}
\begin{equation}
H_R=\lambda (k_y \sigma_x - k_x \sigma_y),
\label{Eq3}
\end{equation}
where $\lambda$ denotes the strength of the RSOC.

The third term in Eq. (\ref{Eq1}), $\Sigma^r$, represents the retarded self-energy originating from the semi-infinite FM lead, which introduces non-Hermiticity into the system through its imaginary components. 
In the wide-band limit \cite{datta1997electronic,bergholtz2019non}, the self-energy term becomes momentum- and frequency-independent and can be expressed as \cite{datta1997electronic,Ryndyk2009}
\begin{equation}
\Sigma^r (\omega=0)= -i \Gamma \sigma_0 -i \gamma \sigma_z,
\label{Eq4}
\end{equation}
where $\Gamma=(\Gamma_\uparrow + \Gamma_\downarrow)/2$ and $\gamma=(\Gamma_\uparrow - \Gamma_\downarrow)/2$ in which $\Gamma_{\uparrow,\downarrow}$ denote the coupling strengths between the UPM and the FM lead, and can be defined as \cite{cayao2023exceptional,reja2024emergence,cayao2022exceptional,cayao2024non}
\begin{equation}
\Gamma_{\uparrow,\downarrow}= \pi \mid t' \mid^2 \rho^{\uparrow,\downarrow},
\label{Eq5}
\end{equation}
where $t'$ and $\rho$ refer to the hopping amplitude into the lead from the UPM and the surface density of states of the lead, respectively. It is worth noting that the self-energy $\Sigma^r$ can possess both real and imaginary components; however, the real part is Hermitian and solely re-normalizes $H$. Conversely, the imaginary part plays a crucial role in NH physics. The detailed derivation of Eq. (\ref{Eq4}) can be found in Ref. \cite{cayao2023exceptional}.

Combining Eqs. (\ref{Eq2}-\ref{Eq4}) gives the full Hamiltonian of the NH proposed system as

\begin{equation}
\begin{split}
    H = &t \left( (\mathbf{k}^2 + \boldsymbol{\alpha}^2) \sigma_0 + 2 \mathbf{k} \cdot \boldsymbol{\alpha} \, \sigma_z \right) - \mu \sigma_0 -i \Gamma \sigma_0 -i \gamma \sigma_z\\
    &+\lambda (k_y \sigma_x - k_x \sigma_y).
\end{split}
\label{Full bare Hamiltonian}
\end{equation}

It is useful to express the Hamiltonian of Eq. (\ref{Full bare Hamiltonian}) in the form $H=\epsilon_0 + \boldsymbol{p} \cdot \boldsymbol{\sigma}$. Here, $\epsilon_0$ is a complex quantity and $\boldsymbol{p}=\boldsymbol{p_r}+i \boldsymbol{p_i}$ with $\boldsymbol{p_r}=\{\lambda k_y, -\lambda k_x, 2 t \alpha_x k_x+ 2 t \alpha_y k_y\}$ and $\boldsymbol{p_i}=\{ 0,0,-\gamma \}$ are the real and imaginary components of the new defined quantity $\boldsymbol{p}$.

The general eigenvalues of Eq. (\ref{Full bare Hamiltonian}) are found to be
\begin{equation}
E_\pm=\epsilon_0 \pm \sqrt{\boldsymbol{p_r}^2-\boldsymbol{p_i}^2+2 i \boldsymbol{p_r} \cdot \boldsymbol{p_i}},
\label{Eq6}
\end{equation}
in which $\epsilon_0=t (\boldsymbol{\alpha}^2+\textbf{k}^2)-i \Gamma -\mu$, leading to
\begin{equation}
\begin{split}
E_\pm = & \ t (\alpha_x^2 + \alpha_y^2 + k_x^2 + k_y^2) - i \Gamma - \mu\\
& \pm \sqrt{(2 t \alpha_x k_x+ 2 t \alpha_y k_y - i \gamma)^2 + \lambda^2(k_x^2 + k_y^2)}.
\end{split}
\label{Eq7}
\end{equation}

The normalized eigenvectors of Eq. (\ref{Full bare Hamiltonian}) also can be written as
\begin{equation}
\psi^\mp=\frac{1}{\beta}
\begin{pmatrix}
\displaystyle \frac{i p + \gamma \mp i \sqrt{(p - i \gamma)^2 + \textbf{k}^2 \lambda^2}}{(k_x + i k_y) \lambda } \\
\\
\displaystyle {1}
\end{pmatrix},
\label{eigenvector}
\end{equation}
where $\beta$ is the normalization factor and $p=2t(k_x \alpha_x+k_y \alpha_y)$.

Unless otherwise stated, wave vectors and UPM strengths are given in units of \( a^{-1} \), and the energies (including the real and imaginary parts of the eigenvalues, the CPL energies ($\Delta$) and magnetic filed ($B_{x,y,z}$) energies) are expressed in units of \( t/ a^2 \), where \( a \) is the lattice constant of the assumed square lattice. The RSOC strength is also given in units of \( t/ a \). The effect of magnetic and optical fields on the Hamiltonian will be discussed in the following sections.

\section{Results and discussion}
Before delving into the EPs, their conditions, and tunability, it is constructive to briefly discuss the Fermi surfaces of UPM, as shown in Fig. \ref{Fig1}(b). Evidently, the Fermi surface in UPM exhibits a circular form, in contrast to the elliptical one in $d$-wave AM. This can be intuitively understood from Eq. (\ref{Eq7}) by setting $\alpha_x=\alpha_y=\Gamma_{\uparrow,\downarrow}=\lambda=0$, which yields a simple circle equation with radius $\sqrt{(E+\mu)/t}$. However, activating $\alpha_x$ ($\alpha_y$) leads to the separation of two Fermi circles corresponding to distinct spin states by $\mid 2\alpha_x \mid$ ($\mid 2\alpha_y \mid$) along the x (y)-direction, namely the $p_x$-wave ($p_y$-wave), as depicted in the left (middle) part of Fig. \ref{Fig1}(b). In the general case, the two Fermi circles are separated by $\boldsymbol{\alpha}=\{\alpha_x,\alpha_y\}$ vector in the ($k_x$-$k_y$) plane; however, their circular form remains unchanged (see the right part of Fig. \ref{Fig1}(b)).
Turning to the case where $\lambda \neq 0$, the RSOC mixes the spin states and induces spin–momentum locking. 
Consequently, the Fermi surfaces become distorted and are no longer circular, which in turn affects the crossing points of the Fermi surfaces (not shown here).

\subsection{Emergence of EPs in UPM-based junction}
Now, we aim to explore the conditions for EP emergence in the proposed FM/UPM junction. Based on Eq. (\ref{Eq6}), the general conditions for EPs can be written as $\boldsymbol{p_r}^2=\boldsymbol{p_i}^2$ and $\boldsymbol{p_r} \cdot \boldsymbol{p_i}=0$ \cite{bergholtz2021exceptional,reja2024emergence,cayao2023exceptional}. These conditions give rise to
\begin{subequations}
\begin{equation}
\lambda^2(k_x^2+k_y^2)+(2 t \alpha_x k_x+2 t \alpha_y k_y)^2=\gamma^2,
\label{Eq8a}
\end{equation}
\begin{equation}
\gamma (2 t \alpha_x k_x + 2 t \alpha_y k_y)=0,
\label{Eq8b}
\end{equation}
\end{subequations}
which gives the non-trivial restriction $(2 t \alpha_x k_x + 2 t \alpha_y k_y)=0$. 
Simultaneous satisfaction of these conditions leads to the emergence of EPs in the proposed FM/UPM junction. This can occur in different scenarios. It must be noted that the two cases of $\gamma=0$ and $\lambda=0$ are excluded from our investigations. The former yields a trivial answer, and the latter is not a reasonable choice due to the presence of the FM lead, which induces the RSOC in the structure. Additionally, and more importantly, neglecting the RSOC eliminates the off-diagonal terms of the Hamiltonian (\ref{Eq1}) and thus annihilates the UPM-dependence of the spinor part of the eigenstates. In such a scenario, achieving non-trivial EPs becomes impossible in this structure. 
The condition $ \lambda^2 (k_x^2 + k_y^2)=\gamma^2 $ in Eq. (\ref{Eq8a}) implies that EPs form a ring in momentum space. As indicated in Ref. \cite{reja2025n}, a N\'eel vector can deform the ring into an elliptical or more complex contours. However, condition (\ref{Eq8b}) in our model, couples the $k_x$ and $k_y$ and thus, restricts the EPs to few discrete points on the ring.

In the following, we examine four representative cases to elucidate how the EP conditions manifest under different UPM configurations ($\alpha_x, \alpha_y$):

\begin{figure}
\includegraphics[scale=0.8]{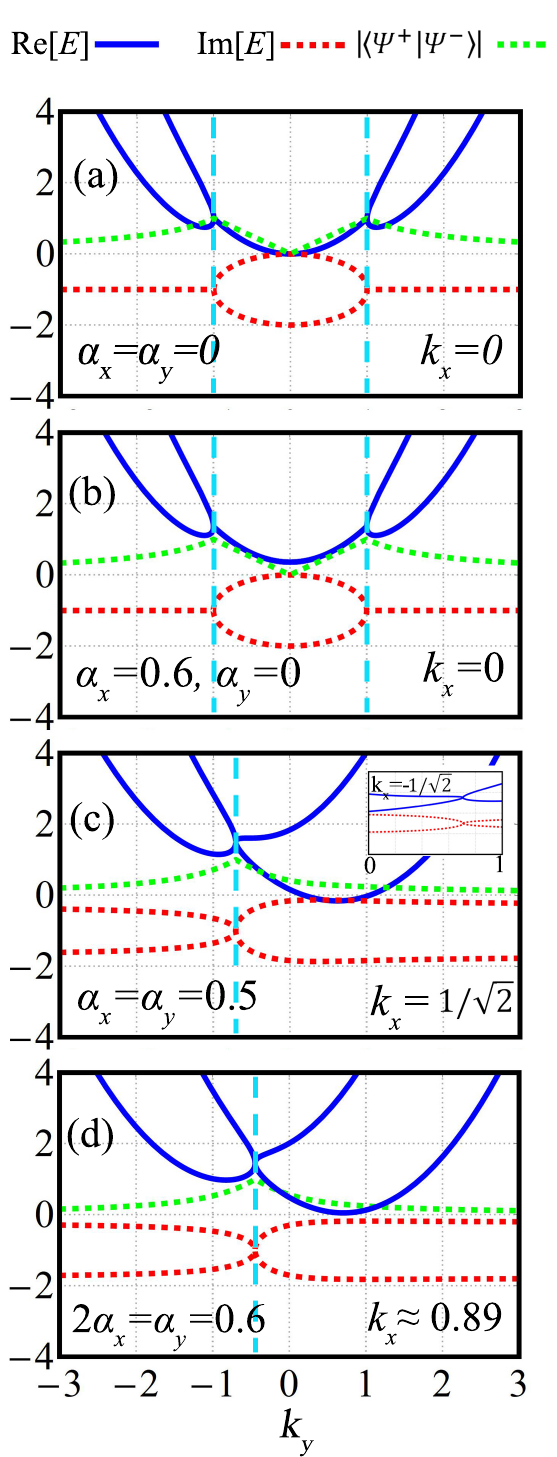}
\caption{ The real and imaginary parts of the energy ($E$) in unit of $t/a^2$ for (a) $\alpha_x=\alpha_y=0$, (b) $\alpha_x=0.6$, $\alpha_y=0$, (c) $\alpha_x=\alpha_y=0.5$, and (d) $2\alpha_x=\alpha_y=0.6$. Here, we have chosen $\lambda=1$, $\Gamma_\uparrow=2$, and $\Gamma_\downarrow=0$. Other fixed parameters are labeled in each panel. The cyan dashed-lines are eye-guides show the approximate $k_y$ of each EP. The green-dashed line indicates the magnitude of overlap of eigenvectors $\psi^+$ and $\psi^-$.} 
\label{Fig2}
\end{figure}

\textbf{Case I}: A familiar form of EPs are those that appear at $k_x=0$ or $k_y=0$ lines. In the proposed structure, setting $k_x=\alpha_x=\alpha_y=0$ simplifies condition (\ref{Eq8a}) to $\lambda^2 k_y^2=\gamma^2$ and also satisfies condition (\ref{Eq8b}). Hence, two EPs appear at $(k_x,k_y)=(0,\pm \gamma/\lambda)$, indicating that EPs can also be found in the absence of UPM, and their positions are tunable via the RSOC strength, which could be adjusted electrically. At $\alpha_x=\alpha_y=0$, the UPM reduces to a Rashba-perturbed semiconductor, as discussed in Ref. \cite{cayao2023exceptional}, confirming our predictions. Notably, based on Eq. (\ref{Eq8a}) EPs can also be observed at $k_y \neq 0$ when $(k_x^2+k_y^2)=(\gamma/\lambda)^2$, which defines a circle of radius $\sqrt{(\gamma/\lambda)^2} = \mid \gamma/\lambda \mid$. 

To further clarify this case, we have plotted the real and imaginary parts of energy ($E$) with respect to $k_y$ for $\boldsymbol{\alpha}=0$ in Fig. \ref{Fig2}(a). Evidently, two points in the ($E-k_y$) plane are observed where the real and imaginary parts for the two spin states merge simultaneously; a manifestation of EPs. These points for $\gamma=\lambda=1$ can be found at $(k_x,k_y)=(0,\pm 1)$. Note that a similar pattern can also be found in the $(E-k_x)$ plane at $k_y=0$ (not shown here). 
The green-dashed line idicates the overlap between the two eigenvectors $\mid \langle\psi^+|\psi^-\rangle \mid$. We observe that at $(k_x,k_y)=(0,\pm 1)$, the two eigenvectors overlap $\mid \langle\psi^+|\psi^-\rangle \mid=1$, confirming the emergence of EPs. Setting $k_x \neq 0$ can also give two EPs if $k_x^2+k_y^2=(\gamma/\lambda)^2$, making a circular pattern \cite{cayao2023exceptional}.

\textbf{Case II}: When the UPM exhibits a $p_x$- or $p_y$-wave nature, i.e., only $\alpha_x$ or $\alpha_y$ is non-zero, respectively, the imaginary part of the energy remains unchanged. Although the real energies are affected by the magnetism of UPM, the positions of the EPs do not change.  All of these effects are shown in Fig. \ref{Fig2}(b), in which the EPs can still be observed at $(k_x,k_y)=(0,\pm 1)$, indicating that the $p$-wave magnetic manipulation cannot affect the EP's positions in this scenario, however, the EPs are no longer circular and only exhibit two discrete points, because $k_x$ should be fixed on zero to satisfy condition (\ref{Eq8b}). The overlap of eigenvectors also demonstrates the persistence of the EP's position (see the green-dashed line in Fig. \ref{Fig2}(b)). These observations can be understood by examining the conditions presented in Eqs. (\ref{Eq8a}) and (\ref{Eq8b}). At $k_x=0$ and $\alpha_y=0$, Eq. (\ref{Eq8b}) is satisfied, regardless of the values of $\alpha_x$. Under this condition, Eq. (\ref{Eq8a}) reduces to a form identical to what is found in previous case, $(k_x^2+k_y^2)=(\gamma/\lambda)^2$. Thus, the main pattern in this case is the same as in Case I. However, applying $\alpha_x \neq 0$ modifies the eigenvalues and re-normalizes the $\textbf{k}^2$ term by increasing the energy, as can be indicated by Eq. (\ref{Eq7}) and seen in the increased real energies of Fig. \ref{Fig2}(b).

\textbf{Case III}: When both $\alpha_x$ and $\alpha_y$ are non-zero (indicating $p_{xy}$-wave of UPM) and equal in magnitude, i.e., $\mid \alpha_x \mid =\mid \alpha_y \mid$, the condition of Eq. (\ref{Eq8b}) is satisfied only at $k_x = -k_y$. This is distinct from the $d_{xy}$-wave in which the EPs only emerge at $k_x k_y=0$ \cite{reja2024emergence,dash2024role}. It is also different from the emergence of EPs at $k_x=\pm k_y$ in the $d_{x^2-y^2}$-wave of AM. In fact, the proposed FM/UPM junction exhibits only two EPs, which are less than the four EPs observed in $d$-wave-based one. This distinction arises from the order of $k_x$ and $k_y$ in their coupling with the UPM term in the eigenvalues of the system, which also can be understood from the nodes in the Fermi surfaces presented in Fig. \ref{Fig1}(a). In the $d$-wave case, this coupling is second order of the wave vector, as $E \propto(k_x k_y)$ for $d_{xy}$ and $E \propto (k_x^2-k_y^2)$ for $d_{x^2-y^2}$ \cite{reja2024emergence}, while in the UPM, it is linear, i.e., $E \propto (2 \alpha_x k_x+2 \alpha_y k_y)$, and its sign also differ from that of $d$-wave, as seen in Eq. (\ref{Eq8b}). This qualitative difference leads to fewer EPs with different conditions in UPM compared to the $d$-wave AM. 

Analytically, Eqs. (\ref{Eq8a}) and (\ref{Eq8b}) shows that the positions of the EPs in the proposed setup are now located at $k_x=-k_y=\pm \gamma/\sqrt{2} \lambda$. The energy states under this condition ($\alpha_x=\alpha_y=0.5$) are plotted in Fig. \ref{Fig2}(c). It is seen that there is an EP at $k_x=-k_y=1/\sqrt{2}$. The second EP is shown in the inset of Fig. \ref{Fig2}(c), indicating the emergence of two EPs at ($k_x,k_y)=(\pm 1/\sqrt{2},\mp 1/\sqrt{2})$. Again, the overlap (dashed-green line) confirms this prediction in Fig. \ref{Fig2}(c). Noting that, the overlap is not shown in the inset to avoid the graphical complexity.

\textbf{Case IV}: The more general case is that the x- and y-components of the magnetization strength in UPM contribute unequally ($\alpha_x \neq \alpha_y \neq 0$). Now, the EPs appear at $\mid k_x \mid \neq \mid k_y \mid$. This feature offers the opportunity to control the direction of EP line (the line which connect two EPs) in the ($k_x$-$k_y$) plane by adjusting the magnetization strength. For $\alpha_x \neq \alpha_y \neq 0$, Eq. (\ref{Eq8b}) leads to $k_x= - (\alpha_y/\alpha_x) k_y $. Combining this with Eq. (\ref{Eq8a}) gives the position of EPs as
\begin{equation}
k_x=\pm \frac{\gamma \alpha_y}{\lambda \sqrt{\alpha_x^2+\alpha_y^2}},
\hspace{5mm}
k_y=\mp \frac{\gamma \alpha_x}{\lambda \sqrt{\alpha_x^2+\alpha_y^2}}.
\label{Eq9}
\end{equation}

Interestingly, these positions yield a tunable rotation angle ($\eta$) of EP line in the ($k_x$-$k_y$) plane of the UPM-based device as
\begin{equation}
\eta= \tan^{-1}(k_y/k_x)= \tan^{-1} (-\alpha_x/\alpha_y).
\label{Eq10}
\end{equation}
This indicates that the orientation of the EP lines in the  ($k_x$-$k_y$) plane depends solely on the magnetization strength and can be effectively modified by these quantities with distinct condition from that of $d$-wave AMs \cite{reja2024emergence,dash2024role}.

The real and imaginary energies versus $k_y$ are plotted in Fig. \ref{Fig2}(d) for $2\alpha_x=\alpha_y=0.6$ at $k_x \approx 0.89$, showing the emergence of an EP under the conditions presented in Eq. (\ref{Eq9}). Another EP emerges at $k_x=-2 k_y \approx -0.89$, but is not shown here. The overlap of eigenstates at this point $\mid \langle\psi^+|\psi^-\rangle \mid=1$ also confirms the occurrence of an EP under this parameter regime.

A comparison of Figs.~\ref{Fig2}(a–d) reveals that, although magnetic control via the UPM layer is not essential for the emergence of EPs, it significantly influences their positions within the FM/UPM junction, without altering their total number. 
It is also evident that the imaginary part of the energy at the EPs remains fixed at $\text{Im}[E] = 1$. This behavior is expected, as the square-root term in the eigenvalue expression~(\ref{Eq7}) vanishes at the EPs, leaving only the base energy $\epsilon_0$, whose imaginary component is set by $-i\Gamma$. Consequently, the imaginary energy at the EPs is directly determined by $\Gamma$. In contrast, the real part of the energy depends on both the wave vector and the UPM magnetization strength, and thus varies across different parameter regimes.

\subsection{Tuning the EPs via UPM properties}

\begin{figure}
\includegraphics[scale=0.6]{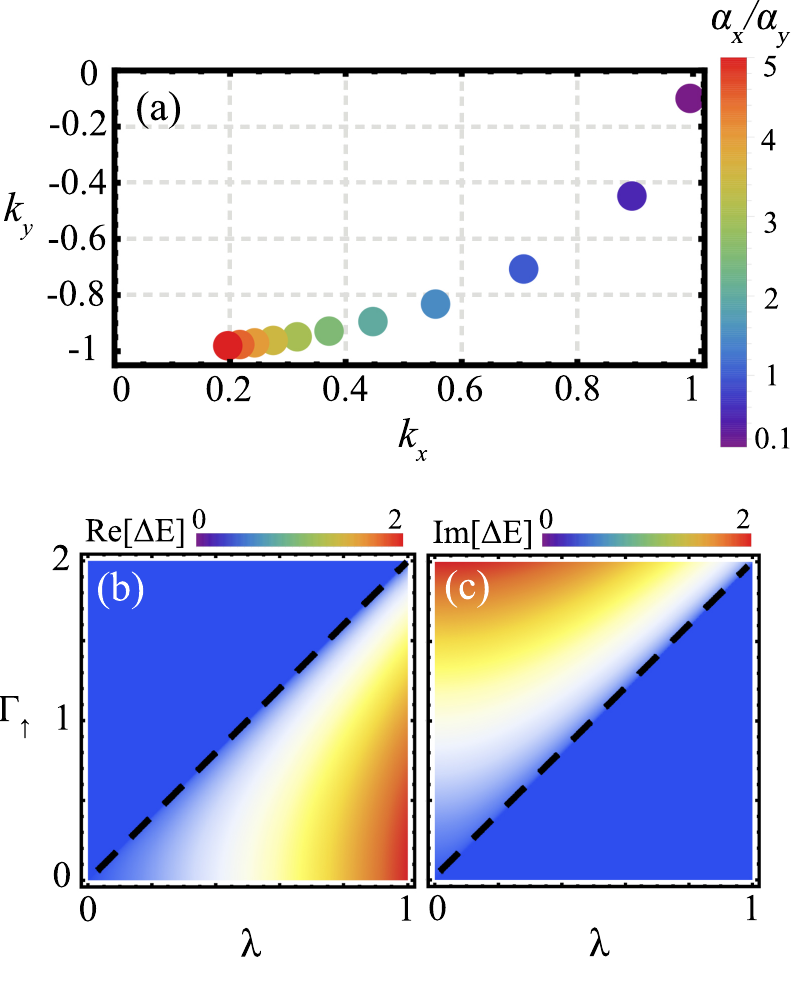}
\caption{ (a) The position of EPs in $(k_x- k_y)$ plane for different ratios of $\alpha_x/\alpha_y$. Density plots of (a) the real part and (b) the imaginary part of the energy difference ($\Delta E= \mid E_+-E_- \mid$) in ($\lambda- \Gamma_\uparrow$) plane for $\Gamma_\downarrow=0$. In (b, c) the fixed parameters are set as $2\alpha_x=\alpha_y=0.6$, $k_x=-2k_y=0.89$, corresponding to the shown EP in Fig. \ref{Fig2}(d). Each colored-point, indicates a certain ratio of ($\alpha_x/\alpha_y$), not an interval.} 
\label{Fig3}
\end{figure}

After finding the possible conditions for realizing EPs in the FM/UPM open junction, we now aim to tune these points. As can be seen in Eq. (\ref{Eq9}), the magnetization strength of the UPM is one of the controlling parameters in this regard. To further clarify this, we have plotted the position of EPs in the ($k_x$-$k_y$) plane for various ratios of ($\alpha_x/\alpha_y$), which are indicated by colors in Fig. \ref{Fig3}(a). Evidently, at low values of ($\alpha_x/\alpha_y$), e.g., 0.1, the EP is approximately located at $(k_x,k_y) \approx (1,-0.1)$. By increasing the ratio of $(\alpha_x/\alpha_y)$, the EP shows a paraboloidal shift towards ($k_x,k_y)\approx (0.2,-1)$, indicating the effective tunability of the EP location by the magnetization strength of UPM. The results for the reversed ratio, i. e., $(\alpha_y/\alpha_x)$, can be found by substituting $k_x \rightarrow -k_y$ and $k_y \rightarrow -k_x$. This means that, for example,  the EP for $(\alpha_x/\alpha_y)=0.1$ can be found at $(k_x,k_y) \approx (1,-0.1)$, while the EP for $(\alpha_y/\alpha_x)=0.1$ is located at $(k_x,k_y) \approx (0.1,-1)$, which is not shown here. Notably, this effect shows that the UPM properties cannot affect the distance of EPs and consequently their number, individually. 

Equation (\ref{Eq9}) also shows the feasibility of controlling EPs through the RSOC strength ($\lambda$) and the FM coupling strength ($\gamma$). To further see how these two factors can tune the EP positions in the proposed junction, the density plots of the real and imaginary parts of the energy difference ($\Delta E= \mid E_+ - E_- \mid$) as a function of $\Gamma_\uparrow$ and $\lambda$ are plotted in Figs. \ref{Fig3}(b) and \ref{Fig3}(c), respectively. Here, $k_x=-2k_y \approx 0.89$ are chosen, corresponding to the EP position in Fig. \ref{Fig2}(d). At first glance, a contrast can be found between the real (Re[$\Delta E$]) and imaginary energy differences (Im[$\Delta E$]), which shows that whenever the Re[$\Delta E$] becomes zero, the Im[$\Delta E$] becomes non-zero, and vice versa. This can also be seen in Figs. \ref{Fig2}(a) and \ref{Fig2}(b) while varying $k_y$, indicating that the EPs appear at certain points, not intervals. Both parameters can control the EP line, i.e., the line in the ($\lambda-\Gamma_\uparrow$) plane where both the real and imaginary parts of the energy merge, as shown by the black-dashed lines. Increasing $\Gamma_\uparrow$ broadens (narrows) the range of $\lambda$ in which the Re[$\Delta E$] (Im[$\Delta E$]) vanishes. The slope of the EP line is determined by the values of \( k_x \) and \( k_y \), which in turn requires appropriate choice of the strength and orientation of the magnetization vector of UPM, the NH coupling, and the RSOC strength. For the chosen parameters (i.e., \( \alpha_y = 2\alpha_x = 0.6 \)), the appearance of EPs requires \( \Gamma_\uparrow > \lambda \).

\subsection{Tuning the EPs via external magnetic field}

\begin{figure*}
\includegraphics[scale=0.8]{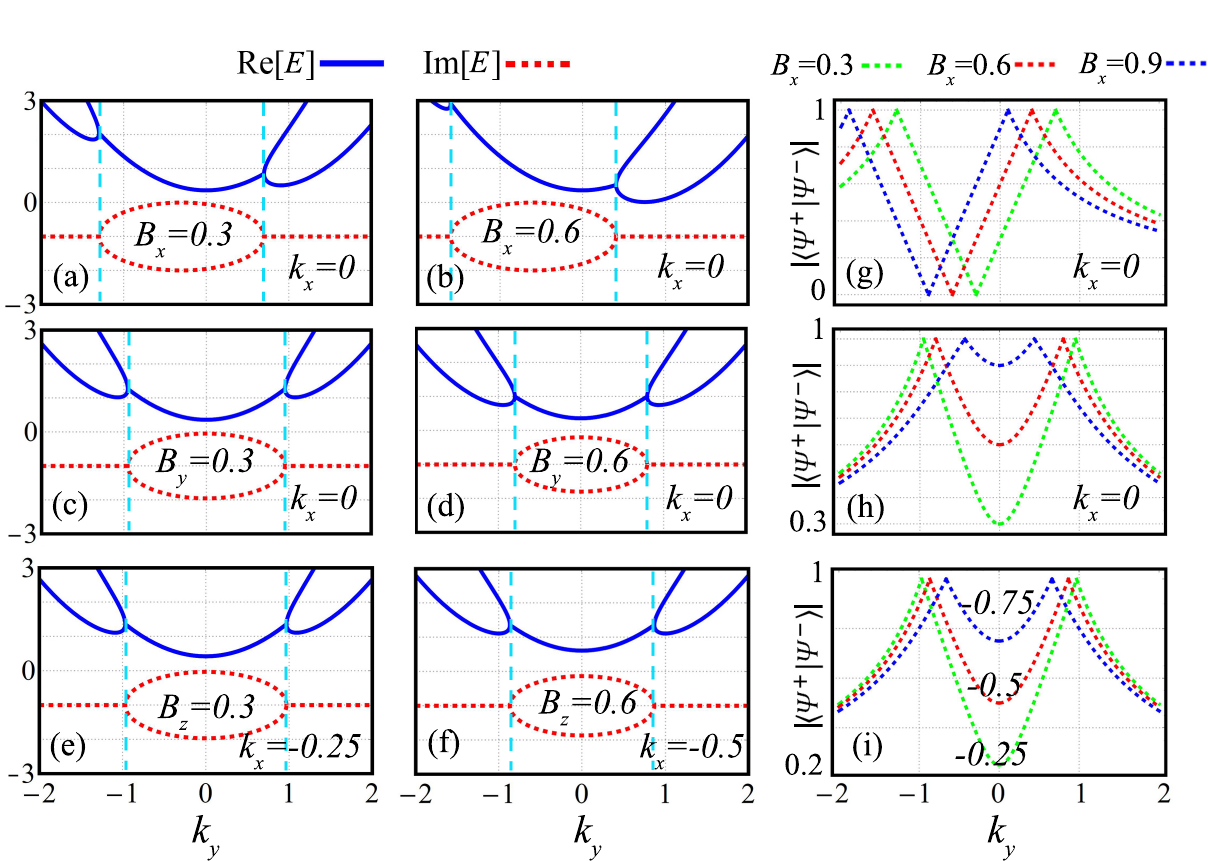}
\caption{ (a-f) The real and imaginary parts of energy with respect to $k_y$ at fixed $k_x$. The first, second, and third rows are plotted in the presence of magnetic filed in x-, y-, and z-direction, respectively. The intensity of $B$ and value of $k_x$ are labeled in each graph. The cyan dashed-lines are eye-guides show the approximate $k_y$ of each EP. (g-i) The overlap of wave functions versus $k_y$ for different magnetic field intensity and directions. In all graphs $\alpha_x=0.6$, $\alpha_y=0$, and other fixed parameters are the same as Fig. \ref{Fig2}. In (i), the green-, red-, and blue-dashed lines indicate $k_x=-0.25, k_x=-0.5$, and $k_x=$-0.75, respectively.  } 
\label{Fig4}
\end{figure*}

After finding the required EP conditions and the tunability of them via magnet vector properties in the proposed FM/UPM junction, we now aim to investigate how an external magnetic field influences the occurrence, number, and position of EPs. 
The Hamiltonian of the proposed system in the presence of a three-dimensional magnetic field can be modeled by adding a term as $H_B=\boldsymbol{B} \cdot \boldsymbol{\sigma}$ to Eq. (\ref{Eq1}) \cite{rao2024tunable,reja2024emergence,dash2024role}, leading to

\begin{equation}
\begin{split}
    H =& t \left( (\mathbf{k}^2 + \boldsymbol{\alpha}^2) \sigma_0 + 2 \mathbf{k} \cdot \boldsymbol{\alpha} \, \sigma_z \right) - \mu \sigma_0+\lambda (k_y \sigma_x - k_x \sigma_y)\\
    &+B_x \sigma_x+B_y \sigma_y+B_z \sigma_z-i \Gamma \sigma_0 -i \gamma \sigma_z.
\end{split}
\label{H with B}
\end{equation}
The presence of the magnetic field extends the energies from what are seen in Eq. (\ref{Eq7}) to
\begin{equation}
\begin{split}
E_\pm = & \ t (\alpha_x^2 + \alpha_y^2 + k_x^2 + k_y^2) - i \Gamma - \mu\\
& \pm \sqrt{B_x^2 + B_y^2+(B_z+ \delta)^2 + \lambda^2 \textbf{k}^2+2 \lambda (B_x k_y-B_y k_x)}.
\end{split}
\label{Eq11}
\end{equation}
where $\delta=2 t \alpha_x k_x+ 2 t \alpha_y k_y - i \gamma$. In the following, we investigate the effect of different directions of $\textbf{B}$ on the position and number of EPs.

$\textbf{Case I}$: $\boldsymbol{\theta=\pi/2$ and $\phi=0}$

In this case, the magnetic field is applied only in the x-direction (i.e., $\textbf{B}=(B_x,0,0)$).
The real and imaginary parts of the energy with respect to $k_y$ at $k_x=0$ and $B_x \neq 0$ are shown in Figs. \ref{Fig4}(a) and \ref{Fig4}(b). According to Eq. (\ref{Eq11}), $B_x$ couples $\lambda$ and $k_y$, alongside re-normalizing the square root. This role shifts the EPs' positions toward negative values of $k_y$. A Comparison of Figs. \ref{Fig4}(a) and \ref{Fig4}(b) shows that increasing the intensity of $B_x$ leads to a more pronounced shift. However, this perturbation cannot change the number of EPs in the ($E,k_y$) diagram. To better see the effect of magnetic field on the EP position and their variation trend, the overlap of eigenvectors $\mid \langle\psi^+|\psi^-\rangle \mid$ is illustrated in Fig. \ref{Fig4}(g) under the influence of $B_x$ for three different strengths. Evidently, both of the unity points (indicating the full overlap) shift toward negative values of $k_y$, indicating the tunability of EP positions via $B_x$ without changing their number.

 Equation (\ref{Eq11}) yields the following two conditions for EP occurrence in the presence of $B_x$
\begin{subequations}
\begin{equation}
(\lambda k_y+B_x)^2+(\lambda k_x )^2+(2 t \alpha_x k_x + 2 t \alpha_y k_y)^2=\gamma^2,
\label{Eq12a}
\end{equation}
\begin{equation}
\gamma(2 t \alpha_x k_x+2 t \alpha_y k_y)=0.
\label{Eq12b}
\end{equation}
\end{subequations}
Evidently, $B_x$ does not change the second condition, because it only couples to off-diagonal $\sigma_x$. 

At $\alpha_y=k_x=0$, the second condition always holds; however, the first one modifies the EP location by $B_x$. These conditions lead to two EPs located at ($k_x,k_y)=(0,\frac{-B_x \pm \gamma}{\lambda})$, which gives $k_y=\{0.7, -1.3\}$ for $B_x=0.3$ and $k_y=\{0.4, -1.6\}$ for $B_x=0.6$, as marked by the cyan dashed-lines in Figs. \ref{Fig4}(a, b), respectively. Obviously, the application of $B_x$ cannot change the distance between EPs, and the EPs always maintain a distance of 2, which is adjustable by the amplitude of $\gamma$ and $\lambda$.

By considering the case in which $\alpha_y \neq 0$ while varying $k_x$ at $k_y=\alpha_x=0$, Eq. (\ref{Eq12b}) remains valid, but Eq. (\ref{Eq12a}) reduces to $B_x^2+\lambda^2 k_x^2= \gamma^2$, which is an even function of $k_x$ and results in EP positions as $(k_x,k_y)=(\pm \sqrt{\gamma^2-B_x^2}/\lambda,0)$, yielding $(k_x,k_y)=(\pm 0.8,0)$ for $B_x=0.6$. Unlike the previous case (i.e., varying $k_y$ at fixed $k_x$), here, the EP distance depends on the $B_x$ magnitude and increasing $B_x$ compress the EPs by shifting them in opposite direction, and no tilt can be found in the real and imaginary parts of the energies. Thus, the $B_x>\gamma$ can annihilate the EPs in ($E-k_x$) plane. This behavior is not shown here, but it is very similar to Fig. \ref{Fig4}(d), which is for $B_y \neq 0$.

Now we can qualitatively discuss the effect of $B_x$ on the EPs when the UPM layer has $p_{xy}$-wave, i.e., $\alpha_x, \alpha_y \neq 0$ (not shown here). Considering $\alpha_x=\alpha_y$ at $\gamma=1$, reduces Eq. (\ref{Eq12b}) to $k_x=-k_y$. Substituting this condition into Eq. (\ref{Eq12a}) gives $k_y=-k_x=\frac{1}{2}(-B_x \pm \sqrt{2-B_x^2})$ at $\lambda=1$. This indicates that the EPs are also tunable by $B_x$ in the presence of $\alpha_y$. Under this condition, $B_x$ can reduce the distance between EPs alongside shifting them. 

In the more general case, by choosing $\alpha_x \neq \alpha_y \neq 0$, Eq. (\ref{Eq12b})
leads to $k_x=-(\alpha_y/\alpha_x)k_y$, resulting in two EPs at

\begin{equation}
k_y=-\frac{\alpha_x^2 B_x \lambda \pm \sqrt{\alpha_x^2[-\alpha_y^2 B_x^2+(\alpha_x^2+\alpha_y^2)\gamma^2)\lambda^2}]}{(\alpha_x^2+\alpha_y^2)\lambda^2}.
\end{equation}

\textbf{Case II}: $\boldsymbol{\theta=\phi=\pi/2}$

In this case, the magnetic field is applied only in the y-direction (i.e., $\textbf{B}=(0,B_y,0)$). Equations (\ref{Eq12b}) and (\ref{Eq8b}) under this condition remains unchanged, but Eq. (\ref{Eq8a}) in the presence of $B_y$ becomes
\begin{equation}
(\lambda k_y)^2+(By-\lambda k_x)^2+ (2 t \alpha_x k_x + 2 t \alpha_y k_y)^2=\gamma^2,
\label{Eq13}
\end{equation}
indicating the tunability of the position of the EPs by $B_y$. However, coupling $B_y$ to $k_x$ leads to a different behavior of $(E,k_y)$ diagram in the presence of $B_y$. To further see this, the real and imaginary parts of the energy are plotted in Figs. \ref{Fig4}(c) and \ref{Fig4}(d). Evidently, the application of $B_y$ does not tilt or shift the energy spectrum in the ($E,k_y$) diagram and instead, it brings two EPs closer. This is due to the elimination of the linear term of $\lambda B_y k_x$ by setting $k_x=0$, which makes the $k_y$ to be an even function of $B_y$. Analytically, EPs at fixed $k_x=0$ and $B_y \neq 0$ are located at $(k_x,k_y)=(0,\pm \sqrt{\gamma^2-B_y^2}/\lambda)$, yielding $(k_x,k_y)\approx ( 0, \pm 0.95)$ for $\alpha_y=0$ and $B_y=0.3$, and also $(k_x, k_y)=(0,\pm 0.8)$ for $B_y=0.6$, as illustrated in Figs. \ref{Fig4}(c) and \ref{Fig4}(d), respectively.

This behavior is confirmed by the eigenvectors overlap, as shown in Fig. \ref{Fig4}(h). We observed that increasing $B_y$ reduces the distance between the two peaks, which are the locations of EPs with respect to $k_y$. Choosing $B_y= \gamma$ merges the EPs to one point and applying $B_y>\gamma$ annihilates them. So, the appropriate tuning of the magnetic field in the FM/UPM junction can also be used to control the number of EPs.

On the other hand, calculating the energies with respect to $k_x$ at fixed $\alpha_x=k_y=0$ to satisfy condition (\ref{Eq12b}), in the presence of $B_y$ leads to two EPs located at ($k_x,k_y)=(\frac{B_y \pm \gamma}{\lambda},0)$, which is not shown here but it is very similar to what is seen in Figs. \ref{Fig4}(a) and \ref{Fig4}(b) in the presence of $B_x$ with only one difference. The distinction is that the application of $B_y$ shifts the energies toward positive values of $k_x$, opposite to what is seen for $B_x$. This crossing behavior between $B_x$ and $B_y$ mainly arises due to their similar appearance in the Hamiltonian with opposite sign. The x- and y-components of the magnetic field are coupled to the x- and y-Pauli matrices, respectively, and hence both are off-diagonal terms that are oppositely coupled to $k_x$ and $k_y$ through the RSOC, shifting the EPs along $k_y$. However, they have opposite signs originating from different signs of RSOC terms, leading to different directions in shifting the energies.

In more general case, i.e., $\alpha_x \neq \alpha_y \neq 0$, the conditions become more complicated. The position of EPs under the influence of $B_y$ at $\gamma=1$ then are located at $k_x=-(\alpha_y/\alpha_x) k_y$ and

\begin{equation}
k_y=\frac{\alpha_x[-\alpha_y^2 B_y \lambda \pm \sqrt{\alpha_y^2(-\alpha_x^2 B_y^2+\alpha_x^2+\alpha_y^2)\lambda^2)}]}{\alpha_y(\alpha_x^2+\alpha_y^2)\lambda^2}.
\end{equation}

\textbf{Case III}: $\boldsymbol{\theta=0}$ regardless of $\boldsymbol{\phi}$

In this case, the magnetic field is applied only in the z-direction (i.e., $\textbf{B}=(0,0,B_z)$). Unlike the two previous cases, $B_z$ couples to $\sigma_z$ and is a diagonal term of the Hamiltonian (\ref{H with B}). Thus, it can modify EPs through two mechanisms, simultaneously. First, it can modify the distance between two EPs, and second, it tunes their location in the ($k_x-k_y$) plane. These features of $B_z$ are shown in Figs. \ref{Fig4}(e) and \ref{Fig4}(f) for $B_z=0.3$ and $B_z=0.6$, respectively. The appropriate value of $k_x$ can in the presence of $B_z$ be determined by $k_x=- B_z/2 t \alpha_x$, yielding $k_x=\{-0.25, -0.5, -0.75 \}$ for $B_z=\{0.3,0.6,0.9\}$ at fixed $\alpha_x=0.6$, respectively. Obviously, in addition to the $k_x$ modification by $B_z$, it also reduces the distance between the two EPs. This modification can be analytically described by these two conditions
\begin{subequations}
\begin{equation}
(2 t \alpha_x k_x + 2 t \alpha_y k_y+B_z)^2+\lambda^2 \textbf{k}^2=\gamma^2,
\label{Eq14a}
\end{equation}
\begin{equation}
\gamma(2 t \alpha_x k_x+2 t \alpha_y k_y+ B_z)=0.
\label{Eq14b}
\end{equation}
\end{subequations}

The first condition indicates the distance reduction between the two EPs while the second one leads to the modification of $k_x$.
 At $\alpha_y=0$, the position of EPs can be determined as
$(k_x, k_y) = ( -B_z/2t\alpha_x,\pm \sqrt{\gamma^2 - (B_z + 2t\alpha_x k_x)^2 - \lambda^2 k_x^2}/\lambda)$. This gives the two EPs at $(k_x,k_y)=(-0.25,\pm 0.96)$ for $B_z=0.3$ and at $(k_x,k_y)=(-0.5,\pm 0.87)$ for $B_z=0.6$, as shown by cyan-dashed lines in Figs. \ref{Fig4}(e) and \ref{Fig4}(f). 

The emergence of EPs and their tunability are confirmed in Fig. \ref{Fig4}(i), which illustrates the overlap of wave functions at different values of $k_x=-0.25,-0.5,-0.75$ for three certain values of $B_z$, to see how the strength and orientation of the magnetic fields affect the EPs. It is evident that increasing $B_z$ reduces the distance between the two EPs, merges them at $B_z=2 \gamma \alpha_x/\lambda$, and finally annihilates them at $B_z>2 \gamma \alpha_x/\lambda$, i.e., $B_z>1.2$ for $\alpha_x=0.6$ and $\gamma=\lambda=1$, where the square root for $k_y$ becomes imaginary; a distinct behavior compared to the FM/$d_{xy}$-wave AM junction \cite{reja2024emergence}.
A Comparison of Figs. \ref{Fig4}(g-i) shows that although the eigenvectors can be perpendicular in the presence of $B_x$, the application of $B_y$ or $B_z$ eliminates the
$\mid \langle\psi^+|\psi^-\rangle \mid=0$ point.

The EPs for $B_z \neq 0$ become more complicated when the UPM has $p_{xy}$-wave form. At $\alpha_x=\alpha_y=\alpha$, condition (\ref{Eq14b}) simplifies to $k_x+k_y+B_z/2t\alpha =0$. Substituting this into Eq. (\ref{Eq14a}) leads to $(k_x,k_y) = (-\frac{B_z}{4 \alpha} \pm \frac{\sqrt{8 \alpha^2 - B_z^2}}{4 \alpha},-\frac{B_z}{4 \alpha} \mp \frac{\sqrt{8 \alpha^2 - B_z^2}}{4 \alpha}$) for $\lambda=\gamma=1$. 
Taking $\alpha_x \neq \alpha_y \neq 0$, the general EP location could be found at $k_x= (-B_z -2 \alpha_y k_y)/2 \alpha_x$ and

\begin{equation}
k_y=- \frac{\alpha_y ^2 B_z \lambda^2 \pm \alpha_x \sqrt{\alpha_y^2 \lambda^2 (4 (\alpha_x^2+\alpha_y^2)\gamma^2- B_z^2 \lambda^2)}}{2 \alpha_y (\alpha_x^2+\alpha_y^2)\lambda^2}.
\end{equation}

\begin{figure}
\includegraphics[scale=0.65]{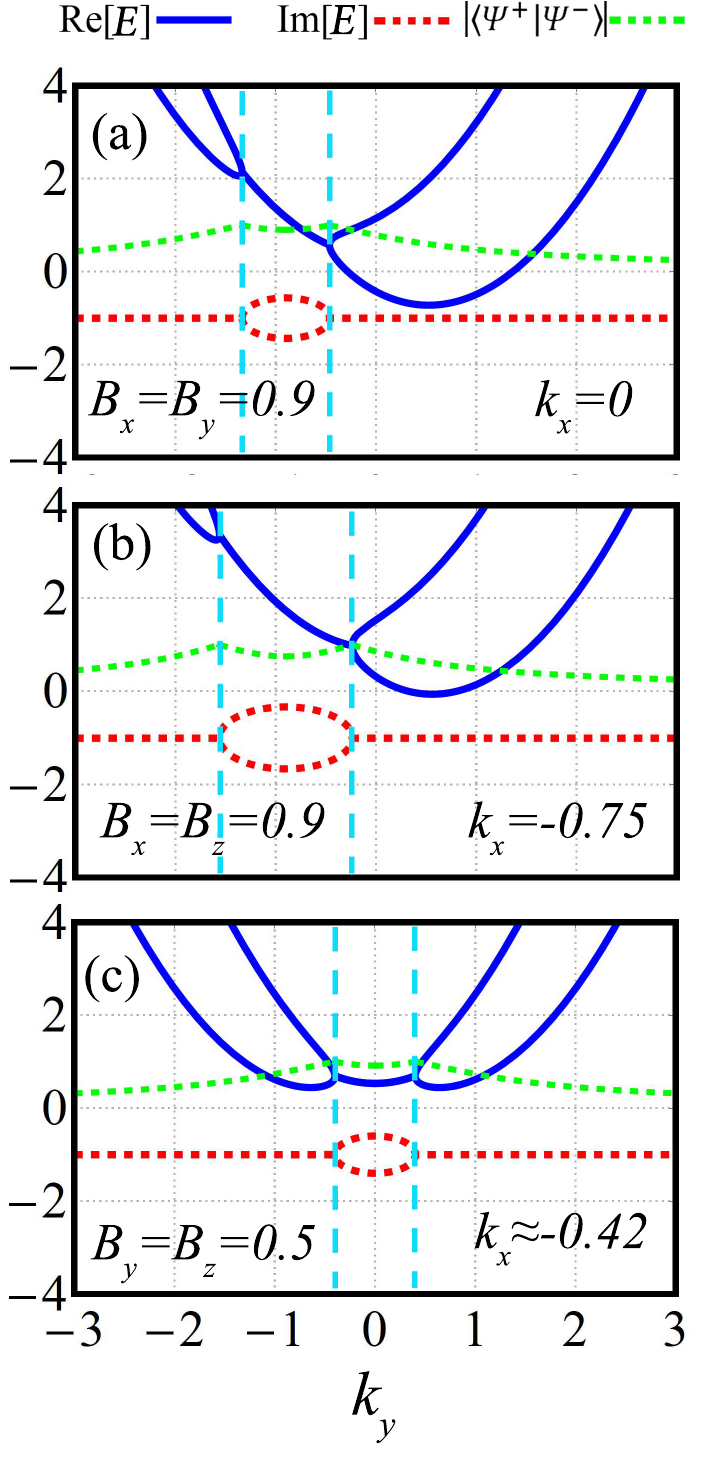}
\caption{ The real and imaginary parts of energy ($E$) with respect to $k_y$ at fixed $k_x$ for (a) $x$-$y$ plane , (b) $x$-$z$ plane, and (c) $y$-$z$ plane magnetic field. In all graphs $\alpha_x=0.6$, $\alpha_y=0$, and other fixed parameters are the same as Fig. \ref{Fig2}, unless it is labeled. The cyan dashed-lines are eye-guides show the approximate $k_y$ of each EP. } 
\label{Fig5}
\end{figure}

\textbf{Case IV}: $\boldsymbol{\theta=\pi/2 ,\phi \neq \{0,\pi/2,\pi,3\pi/2 \}}$

From now on, we investigate the cases with planar magnetic fields. It must be noted that using intermediate angles of $\phi$ (e.g. $\phi=\pi/3$) in planar magnetic fields causes an additional numerical factor due to the $\cos(\phi) \neq 0$, however, we have neglected this factor to simplify the analysis.
In the first case, the magnetic field is applied in the ($x-y$) plane, i.e., $\textbf{B}=(B_x,B_y,0)$. This case is shown in Fig. \ref{Fig5}(a) for $B_x=B_y=0.9$ in the presence of $\alpha_x=0.6$. As expected from Figs. \ref{Fig4}(a)-\ref{Fig4}(d), the simultaneous application of $B_x$ and $B_y$ shifts the EPs alongside reducing their distance in the ($E,k_y$) diagram, and tilt the real energy curves. In this case, the EPs can still be found at $k_x=0$. Interestingly, the application of $B_x$ does not change the allowed interval of $B_y$, and applying $B_y>1$ still annihilates the EPs. The condition presented in Eq. (\ref {Eq12b}) still holds in this case; however, the other required condition now becomes a combination of Eqs. (\ref{Eq12a}) and (\ref{Eq13}), leading to
\begin{equation}
(\lambda k_y+Bx)^2+(By-\lambda k_x)^2+(2 t \alpha_x k_x + 2 t \alpha_y k_y)^2=\gamma^2,
\label{Eq15}
\end{equation}
which, by setting $\alpha_y=0$, leads to two EPs at ($k_x,k_y$)=($0,-B_x/ \lambda \pm \sqrt{\gamma^2  - B_y^2 }$/$\lambda)$ resulting in two EPs at ($k_x,k_y$) $\approx$ (0, -0.46) and ($k_x,k_y$) $\approx$ (0, -1.34) for $B_x=B_y=0.9$ (assuming $\gamma=\lambda=1$), as shown in Fig. \ref{Fig5}(a). This indicates the independence of $\alpha_x$ for the EP position under these specific conditions. Notably, plotting the energies with respect to $k_x$ at $k_y=\alpha_x=0$ and $\alpha_y =0.6$ gives the EPs at ($k_x,k_y$)=(0.46, 0) and ($k_x,k_y$)=(1.34, 0), indicating opposite shifting of EPs (not shown here).

For the $p_{xy}$-wave UPM with $\alpha_x \neq \alpha_y$, condition (\ref{Eq12b}) limits the choice of $k_x$ to $-(\alpha_y/\alpha_x) k_y$. Substituting this condition into Eq. ($\ref{Eq15}$), leads to
\begin{equation}
\begin{split}
k_y=&\frac{-\alpha_x(\alpha_x B_x+\alpha_y B_y)}{\lambda(\alpha_x^2+\alpha_y^2)}\pm\\&\frac{\alpha_x^2\sqrt{[(-(\alpha_y B_x-\alpha_x B_y)^2+(\alpha_x^2+\alpha_y^2)\gamma^2)\lambda^2]/\alpha_x^2}}{\lambda^2(\alpha_x^2+\alpha_y^2)},
\end{split}
\label{Eq16}
\end{equation}
which approximately gives two EPs at ($k_x,k_y)\approx(0.13,-0.26$) and ($k_x,k_y)\approx(0.95,-1.89$), for $2\alpha_y=\alpha_x=0.6$ and $B_x=B_y=0.9$ (not shown here). Obviously, now the EP position relies on the $\alpha_x$ and $\alpha_y$.

\textbf{Case V}: $\boldsymbol{\theta\neq \{0,\pi/2\}$ and $\phi=0}$

In this case, the magnetic field is applied only in the ($x$-$z$) plane (i.e., $\textbf{B}=(B_x,0,B_z)$).The EPs now appear at $k_x = -B_z/2 t \alpha_x$ for $\alpha_y=0$, regarding condition (\ref{Eq14b}). Additionally, the EP occurrence needs to satisfy
\begin{equation}
B_x^2+(2 t \alpha_x k_x + 2 t \alpha_y k_y+B_z)^2+\lambda^2 \textbf{k}^2+ 2 \lambda B_x k_y=\gamma^2.
\label{Eq17}
\end{equation}

A negative shift and an energy tilt can again be observed regarding the application of $B_x$. Additionally, applying $B_z \neq 0$ reduces the distance between the two EPs. Thus, applying both of these fields simultaneously exhibits both aforementioned behaviors. Interestingly, the distance reduction due to $B_z$ is weaker than that of $B_y$ seen in Fig. \ref{Fig5}(a), which can be attributed to the non-zero $k_x$ acquired by the application of $B_z \neq 0$. The combination of Eqs. (\ref{Eq17}) and (\ref{Eq14b}) gives the location of the EPs for $\alpha_y=0$ as $(k_x,k_y)=(-B_z/2 t \alpha_x,-B_x /\lambda \pm \sqrt{4 \alpha_x^4\gamma^2 \lambda^2 - \alpha_x^2 B_z^2 \lambda^4}/2 \alpha_x^2 \lambda^2)$, which at $\alpha_x=0.6$ and $B_x=B_z=0.9$ gives ($k_x,k_y)\approx(-0.75,-0.24$) and ($k_x,k_y)\approx(-0.75,-1.56$).

Generally, setting arbitrary magnetude of magnetic fields in the $x$- and $z$-directions for the $p_{xy}$-wave UPM-based setup leads to $k_x= (-B_z -2 \alpha_y k_y)/2 \alpha_x$ and
\begin{widetext}
\begin{equation}
k_y=-\frac{2 \alpha_x^2 \lambda B_x+\alpha_y B_z \lambda^2 \pm \sqrt{\alpha_x^2 \lambda^2 [4 \alpha_x^2 \gamma^2 -(2 \alpha_y B_x-2 \alpha_y \gamma- B_z \gamma)(2 \alpha_y (B_x+\gamma)-B_z \lambda)]}}{2 \lambda^2(\alpha_x^2+\alpha_y^2)}.
\label{Eq18b}
\end{equation}
\end{widetext}
This indicates the effect of the manipulation of UPM properties on the EP location in this case.

$\textbf{Case VI}$: $\boldsymbol{\theta\neq {0,\pi/2,\pi}$ and $\phi=\pi/2}$

In the final case, the magnetic field is applied in the ($y$-$z$) plane (i.e, $\textbf{B}=(0,B_y,B_z)$. It was shown in Figs. \ref{Fig4}(c)- \ref{Fig4}(f) that both $B_y$ and $B_z$, reduce the distance between EPs and thus, can annihilate them. Hence, when both $B_y$ and $B_z$ are applied while varying $k_y$, the maximum value of $\boldsymbol{B}$ should be restricted carefully. To ensure the real value of $k_y$ avoiding EP anihillation, for $B_z=B_y$, they should satisfy $B_{y,z}< 2 \sqrt{\alpha_x^2/(1+2 \alpha_x)^2}$. For $\alpha_x=0.6$ this restricts $B_z$ and $B_y$ up to 0.54. Thus, Fig. \ref{Fig5}(c), shows the real and imaginary energies for $B_z=B_y=0.5$, which is less than what is chosen in Figs. \ref{Fig5}(a) and \ref{Fig4}(b). This set of parameters at $\alpha_y=0$, leads to two EPs at $(k_x,k_y)=(-B_z/2 \alpha_x, \pm \sqrt{4 \alpha_x^2 - 4 \alpha_x^2 B_y^2-4 \alpha_x B_y B_z-B_z^2}/ 2\alpha_x)$ for $\gamma=\lambda=1$,that corresponds to $(k_x,k_y)\approx(-0.42,\pm 0.40)$ for the chosen parameters of Fig. \ref{Fig5}(c) and obeys

\begin{equation}
B_y^2+(2 t \alpha_x k_x + 2 t \alpha_y k_y+B_z)^2+\lambda^2 \textbf{k}^2- 2 \lambda B_y k_x=\gamma^2,
\label{Eq19}
\end{equation}
and also Eq. (\ref{Eq14b}). 
Generally, the critical value of $B_y$ at $\alpha_y=0$ can be found as $B_y=\gamma - (B_z \lambda/ 2 \alpha_x)$. This relation shows a competition between $B_y$ and $B_z$ in the EP distance variation, in the way that applying $B_y$ after this critical value (that depends on $B_z$) annihilates the EPs.

In a more general situation in which $\alpha_{y,x} \neq 0$, two EPs can be found at

\begin{subequations}
\begin{equation}
k_x=-\frac{-2\alpha_y^2 B_y \lambda+\alpha_x B_z \lambda^2\pm \sqrt{\chi}}{2 \lambda^2(\alpha_x^2+\alpha_y^2)},
\label{Eq20a}
\end{equation}
\begin{equation}
k_y=\frac{-\alpha_y^2 \lambda(2 \alpha_x B_y+\lambda B_z) \pm \alpha_x \sqrt{\chi}}{2 \lambda^2 \alpha_y(\alpha_x^2+\alpha_y^2)},
\label{Eq20b}
\end{equation}
\end{subequations}
  with
  \begin{equation}
  \chi=\alpha_y^2 \lambda^2(4 \alpha_y^2 \gamma^2+4 \alpha_x^2(\gamma^2-B_y^2)-4 \alpha_x B_y B_z \lambda-B_z^2 \lambda^2).
  \end{equation}
 
 \subsection{Optical tuning of the EPs}

\begin{figure*}
\includegraphics[scale=0.8]{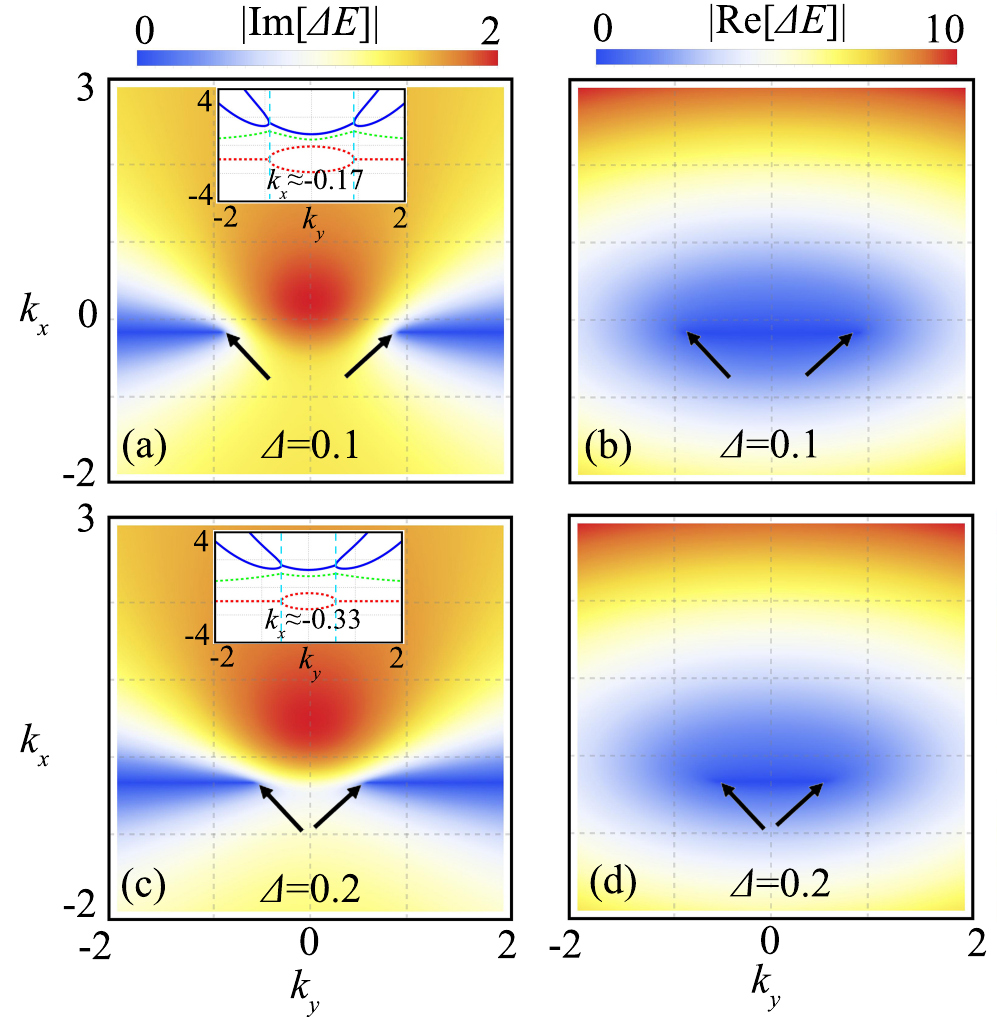}
\caption{The density plots of (a, c) imaginary and (b, d) real parts of the absolute value of energy differences as a function of ($k_x, k_y$). We have fixed $\Delta=0.1$ for (a, b) and $\Delta=0.2$ for (c, d). The insets of (a) and (c) show the linear plot of real (blue) and imaginary (dashed red) at fixed $k_x$ with respect to $k_y$. In all graphs, $\Gamma=\gamma=\lambda=1$, $\alpha_x=0.6$, and $\alpha_y=0$. The black arrows show the approximate location of EPs. We have fixed $\Omega=2$ to ensure the regime required for the high-frequency Floquet framework as used in Ref. \cite{yarmohammadi2025anisotropic}}
\label{Fig7}
\end{figure*}

Although static methods like applying a Zeeman field have proven useful to control the spin-dependent properties of even-parity AMs  \cite{amundsen2024rkky}, dynamic techniques, particularly Floquet engineering, provide a more flexible way to alter the electronic band structure. In our study, we focus specifically on using CPL in off-resonance regime. This choice is deliberate because linearly polarized light only shifts the system’s overall energy, which is not relevant to our goals.

The right-handed CPL can be modeled as a time-dependent vector
potential, given by $\textbf{A}(t)= A [ \sin(\Omega t), \cos(\Omega t)]$, where $A = E_0/\sqrt{2}\Omega$, with $E_0$ and $\Omega$ representing the amplitude and frequency of the light, respectively. This vector potential couples to the Hamiltonian through
the minimal coupling approach (substituting $\textbf{k}$ with $\textbf{k}-e \textbf{A}(t))$, where $e$
is the electron’s charge), inducing transitions between the system’s eigenstates. Given the time-periodic nature of the Hamiltonian, we employ the Floquet-Bloch theorem \cite{kitagawa2011transport,platero2004photon} to address such time-dependent systems. 

Considering the CPL to be in high-frequency regime (larger than the bandwidth), the Floquet sidebands are sufficiently well-separated and the effective Floquet Hamiltonian can be derived using a perturbative technique known as the van Vleck expansion, leading to \cite{yarmohammadi2025anisotropic,kitagawa2011transport,ezawa2013photoinduced}


\begin{equation}
H^F(\textbf{k}) \approx H_0^F+\frac{[H_{-1}^F,H_{+1}^F]}{\Omega},
\end{equation}
in which $H_0^F=1/T \int_{0}^{T} H (\textbf{k}-e\textbf{A}) dt$ and $H_{\pm1}^F=1/T \int_{0}^{T} H (\textbf{k}-e\textbf{A}) e^{\pm i \Omega t} dt$. Here, the result is restricted up to first order. Defining $\Delta=(e A \lambda)^2/2 \Omega$ and evaluating the above integrals leads to the effective Hamiltonian for the UPM in the presence of RSOC as (see appendix \ref{AppA})

\begin{equation}
\begin{split}
H^F \approx & \frac{\hbar^2}{2m}[\textbf{k}^2+\boldsymbol{\alpha}^2+ \frac{2 \Omega \Delta}{\lambda^2}]\sigma_0+ [\frac{\hbar^2}{m}(\textbf{k} \cdot \boldsymbol{\alpha})+2 \Delta] \sigma_z \\
&+[\lambda k_y-\frac{2 \alpha_y \hbar^2 \Delta}{m \lambda}]\sigma_x-[\lambda k_x-\frac{2 \alpha_x \hbar^2 \Delta}{m \lambda}]\sigma_y.
\end{split}
\label{Eq22}
\end{equation}

The eigenvalues of $H^F$ are given by 
\begin{equation}
\begin{split}
  E_\pm=&\boldsymbol{\alpha^2}+\textbf{k}^2-i \Gamma -\mu+(2 \Delta \Omega)/\lambda^2 \pm \\
   & \sqrt{16 \boldsymbol{\alpha^2} \Delta^2 \lambda^2+ [\delta^2- 4 i \gamma \Delta+4 \Delta^2]\lambda^4+\textbf{k}^2 \lambda^6}/\lambda^2.
    \end{split}
\end{equation}

A Comparison between Hamiltonians ~(\ref{H with B}) and~(\ref{Eq22}) shows that while both CPL and $B_z$ reduce the EP separation and modify the $k_x$ term, their mechanism is qualitatively different. The $B_z$ couples only to $\sigma_z$ and leaves RSOC (off-diagonal) terms unaffected. In contrast, thanks to the RSOC, CPL couples to all spin Pauli matrices, leading to UPM-dependent renormalization of RSOC, as seen in Eq.~(\ref{Eq22}). Unlike its effect in $d$-wave AMs~\cite{yarmohammadi2025anisotropic}, CPL in UPM does not couple directly to momentum, but instead modifies the $\lambda k_{x,y}$ terms, which can be attributted to the odd parity of the UPM.

By adding the NH effects as Eq. (\ref{Eq1}) to the Hamiltonian of Eq. (\ref{Eq22}), the conditions for EP occurrence can be given as

\begin{subequations}
\begin{equation}
\begin{split}
&(\lambda k_y-\frac{2 \alpha_y \hbar^2 \Delta}{m \lambda})^2+(-\lambda k_x+\frac{2 \alpha_x \hbar^2 \Delta}{m \lambda})^2\\
&+(\frac{\hbar^2}{m}(\textbf{k} \cdot \boldsymbol{\alpha})+2 \Delta)^2=\gamma^2,
\label{Eq23a}
\end{split}
\end{equation} 

\begin{equation}
\gamma( \alpha_x k_x + \alpha_y k_y+ \Delta)=0.
\label{Eq23b}
\end{equation}
\end{subequations}

Setting $\Delta=0$ in these conditions, simplifies them to Eqs. (\ref{Eq8a}) and (\ref{Eq8b}). Noting that, when $\lambda << 1$, the CPL effect in $2 \Delta \sigma_z$ term of Eq. (\ref{Eq22}) becomes negligible.

The condition presented in Eq. (\ref{Eq23b}), restricts the $k_x$ to $-(\alpha_y+\Delta)/\alpha_x$. Using this limit, the EPs for $\alpha_y=0$ can be found at $k_x=-\Delta/\alpha_x$ and

\begin{equation}
k_y=\pm\frac{\sqrt{-(4 \alpha_x^2 \Delta-\alpha_x \gamma \lambda+\Delta \lambda^2)(4 \alpha_x^2 \Delta+\alpha_x \gamma \lambda+\Delta \lambda^2)}}{\alpha_x \lambda^2}.
\label{Eq24}
\end{equation}
This leads to two EPs at $(k_x,k_y) \approx( -0.17,\pm 0.91)$ for $\alpha_x=6 \Delta=0.6$ and at $(k_x,k_y) \approx (-0.33,\pm 0.58)$ for $\alpha_x=3 \Delta=0.6$. 
To clarify this, Fig. \ref{Fig7}(a) and \ref{Fig7}(b) illustrate the density plots of imaginary and real energy differences, respectively, in the ($k_x-k_y$) plane at $\Delta=0.1$. Evidently, there are two points in which the real and imaginary energies merge (i.e., their difference vanishes), simultaneously. These two are marked by the black arrows in Fig. \ref{Fig7}. Interestingly, applying right-handed CPL affects both the position and separation of the EPs in the ($k_x- k_y$) plane, such that increasing $\Delta$ reduces the distance between two EPs and also moves them toward more negative values of $k_x$. This also can be understood from the insets in Fig. \ref{Fig7}(a) and \ref{Fig7}(c) in which the EPs become closer by increasing $\Delta$ from 0.1 to 0.2. Importantly, Eq. (\ref{Eq24}) indicates that the CPL can reduce the number of EPs and annihilates them beyond $\Delta>(\alpha_x \gamma \lambda)/(4\alpha_x^2+ \lambda^2)$. Numerically, applying $\Delta>0.25$ annihilates the EPs (for the fixed values considered in Fig. \ref{Fig7}). 
Notably, substituting $\alpha_x \neq 0$ with $\alpha_y \neq 0$, can give similar behaviors for $(k_y-k_x)$ plane, i.e., rotate the graph in Fig. \ref{Fig7}(a) by 90 degrees, in the way that the two EPs appear at fixed $k_y$ and different $k_x$ values. 

We now consider the more general case where both $\alpha_x$ and $\alpha_y$ are non-zero. As indicated by Eq.~(\ref{Eq10}), the simultaneous presence of $\alpha_x$ and $\alpha_y$ causes a rotation of the EP line. This results in the EPs forming a diagonal configuration, rather than being aligned along a principal axis. 
 Now, satisfying Eqs. (\ref{Eq23a}) and (\ref{Eq23b}), gives the position of EPs at 
\begin{subequations}
\begin{equation}
k_x = - \frac{
\alpha_x \, \Delta \, \lambda^4 \pm 
\sqrt{\nu
}
}{
\boldsymbol{\alpha}^2 \lambda^4
},
\label{Eq25a}
\end{equation}
\begin{equation}
k_y =  \frac{
-\alpha_y^2 \, \Delta \, \lambda^4 \pm 
\alpha_x \sqrt{\nu
}
}{
\alpha_y \boldsymbol{\alpha^2} \lambda^4
},
\label{Eq25b}
\end{equation}
\end{subequations}
$\nu=\alpha_y^2 \, \lambda^4 [ 
-16 \boldsymbol{\alpha}^2 \Delta^2 + 
\boldsymbol{\alpha}^2(\gamma^2 - 8 \Delta^2) \lambda^2 - 
\Delta^2 \lambda^4]$.


Considering $\alpha_x=3 \alpha_y=0.6$, gives two EPs located at ($k_x,k_y$)=(-0.48,0.44) and ($k_x,k_y$)=(-0.12,-0.64), that are not shown here. Equations (\ref{Eq25a}) and (\ref{Eq25b}) also show that the position of EPs under influence of the CPL highly depends on the RSOC strength, which can be attributed to the role of CPL in Eq. (\ref{Eq22}) in the modification of the $\lambda$ terms. So, the EPs in the proposed device are highly tunable via CPL.

When the irradiated CPL is changed from right-handed to left-handed (substituting red arrows with blue ones in Fig. \ref{Fig1}(a)),  $\Delta$ should be replaced with $-\Delta$ in Eq.~(\ref{Eq22}). This substitution leads to the following changes in the energy difference profile in the $(k_x- k_y)$ plane:

\textbf{Case I}. For $\alpha_x \neq 0$ and $\alpha_y = 0$, replacing $\Delta \rightarrow -\Delta$ reflects the graph across the line $k_x = 0$, i.e., $\Delta E(\Delta, k_x, k_y) = \Delta E(-\Delta, -k_x, k_y)$. This symmetry appears in both the real and imaginary parts of the energy difference. The origin of this behavior can be understood analytically from conditions~(\ref{Eq23a}) and~(\ref{Eq23b}). Setting $\alpha_y = 0$ in these expressions yields Eq.~(\ref{Eq24}), which is odd in $k_x$ with respect to $\Delta$, but even in $k_y$ with respect to $\Delta$. Therefore, changing the sign of $\Delta$ only affects the $k_x$ dependence.

\textbf{Case II}. When $\alpha_y \neq 0$ and $\alpha_x = 0$, the energy difference diagrams rotate by 90 degrees. Then flipping the sign of $\Delta$ results in a reflection of the energy spectrum across the line $k_y = 0$. In other words, the relation $\Delta E(\Delta, k_x, k_y) = \Delta E(-\Delta, k_x, -k_y)$ holds. This symmetry can also be confirmed analytically by substituting $\alpha_x = 0$ into Eqs.~(\ref{Eq23a}) and (\ref{Eq23b}), which yields an energy difference that is odd in $k_y$ and even in $k_x$ under the transformation $\Delta \rightarrow -\Delta$. 

\textbf{Case III}. Considering a more general case in which both $\alpha_x$ and $\alpha_y$ are non zero, leads to two EPs introduced by Eqs. (\ref{Eq25a}) and (\ref{Eq25b}). Evidently, the first term of both $k_x$ and $k_y$ are now proportional to the first order of $\Delta$, while the square root only depends on $\Delta^2$ and does not change under the $\Delta \rightarrow -\Delta$ substitution. 
As an example, one EP for right-handed CPL with $\Delta=0.2$ for $\alpha_x=3\alpha_y=0.6$ is located at ($k_x,k_y$)=(-0.48,0.44), while this EP under influence of left-handed CPL can be found at $(k_x,k_y)=(0.12,0.64)$.

 \subsection{Spin projections}

After demonstrating that the EPs can occur within the RSOC-perturbed FM/UPM junction, we now shift our focus to examining spin behavior in the presence of NH effects.
Specifically, our attention here is on the expectation values of spin (referred to in this section as spin projections) which are calculated using the following expression \cite{cayao2023exceptional}:
\begin{equation}
S_{x(y,z)}^\pm= \psi_\pm^\dag \sigma_{x(y,z)} \psi_\pm,
\label{Eq27}
\end{equation}
in which $\psi_\pm$ are the non-normalized eigenstates of Hamiltonian (\ref{Eq1}). Noting that, here, we focus on the in-plane spin projections ($S_{x,y}^\pm$). Equation (\ref{Eq27}) can be simplified in different situations; however, the general solution is complicated. For $\alpha_x=\alpha_y=\Gamma_{\uparrow,\downarrow}=0$, which corresponds to a Hermitian Rashba semiconductor/FM junction, $S_{x(y)}$ becomes proportional to $\pm k_{y(x)}/\mid \textbf{k} \mid$, similar to what is reported in Refs. \cite{chen2021spin,cayao2023exceptional}. Under this situation, the spin projection undergoes a sign shift while varying the momenta and diverges near $\mid \textbf{k} \mid=0$. Adding non-Hermiticity ($\gamma \neq 0$) re-normalizes the momentum-dependence of $S^\pm$, leading to imaginary values for spin projections, which can be physically interpreted as a signal of their lifetime \cite{cayao2023exceptional}. The emergence of non-Hermiticity also leads to the coalescence of spin projections at EPs, which can be used as a verification tool for EPs and their behavior, as we are looking for here.
 
  \begin{figure}
\includegraphics[scale=0.6]{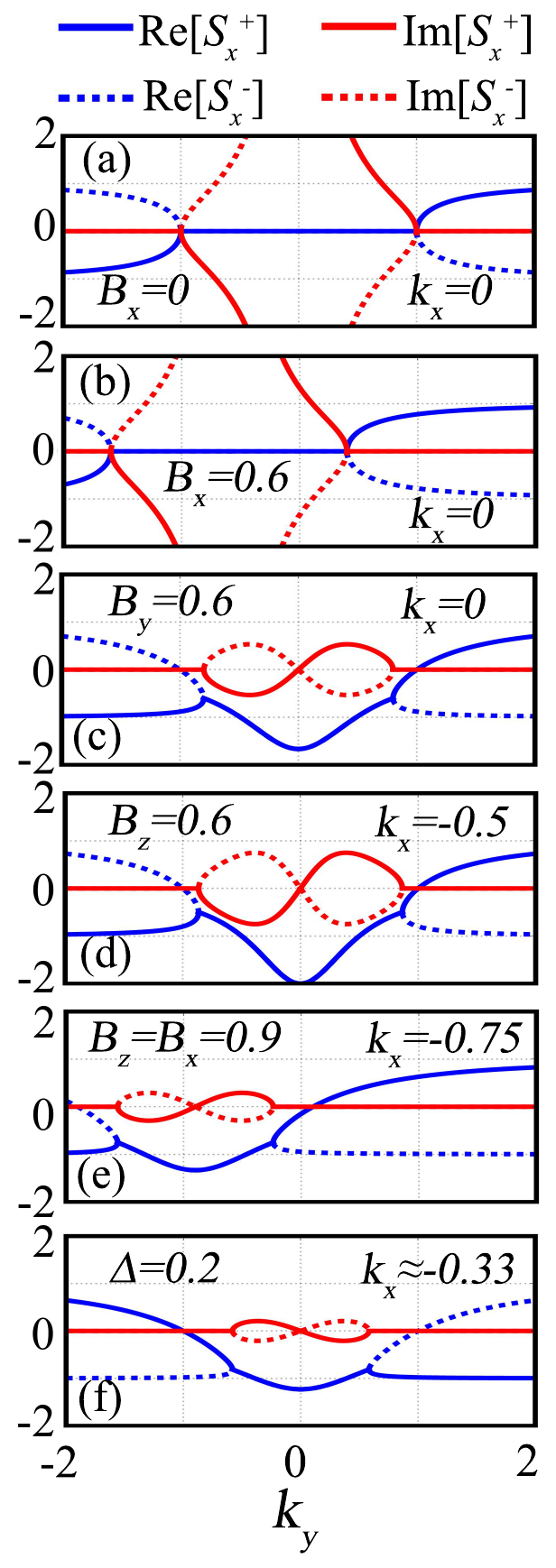}
\caption{ The real and imaginary parts of spin projections with respect to $k_y$ for (a) $B_x=0$, (b) $B_x=0.6$, (c) $B_y=0.6$, (d) $B_z=0.6$, (e) $B_x=B_z=0.9$, and (f) $\Delta=0.2$. The wave vector is labeled in each panel and other fixed parameters are the same as Fig. (\ref{Fig2}).} 
\label{Fig6}
\end{figure}

When the UPM is added, $S^\pm$ becomes more complicated. Now, the $S_x^\pm$ at $\alpha_y=k_x=0$ follows the trend of $\pm2\sqrt{-\gamma^2-k_y^2 \lambda^2}/k_y \lambda$. To further explore this, the $S^\pm_x$ with respect to $k_y$ is plotted in Fig. \ref{Fig6} for different parameter regimes. Noting that, in this figure the calculated spin projections have been uniformly scaled by a factor of $1/2$ for enhancing visual comparison. As seen in Fig. \ref{Fig6}(a), in the absence of magnetic fields ($B=0$), at $k_x=0$, the real parts of $S_x^\pm$ show a sign change while varying $k_y$ after a zero profile. At the end of this zero profile, the real parts of $S_x^\pm$ converge. Between these two degenerate points and when the real part vanishes, the imaginary part of $S_x^\pm$ becomes non-zero and diverges at $k_y \rightarrow 0$. Evidently, the merging of the real and imaginary parts of $S_x^\pm$ is located at the same momenta where the EPs can be seen (see Fig. \ref{Fig2}(b)). Figure \ref{Fig6}(b) shows that turning on the magnetic field in the $x$-direction ($B_x \neq 0$) shifts the convergence points in the spin projections to the left, as same as the EP shifting in the energy spectrum presented in Fig. \ref{Fig4}(b). The divergence of $S_x^\pm$ still can be seen, however, the corresponding $k_y$ also shifts to left.

Despite $B_x$, $B_y$ can modify the distance between EPs while varying $k_y$. As seen in Fig. \ref{Fig6}(c), which illustrates the real and imaginary parts of $S_x^\pm$ versus $k_y$ at $B_y=0.6$, the coalescence effect of the real and imaginary spin projections persists. In this case, the real part does not vanish anymore and instead develops a minimum at $k_y=0$ favoring a large negative spin projection along $x$. Interestingly, the Im[$S^\pm$] does not diverge anymore. This behavior, alongside the non-zero Re[$S^\pm$], arises from the contribution of $B_y$ in the denominator of $S_x^\pm$ due to the $\sigma_y$ coupling in the Hamiltonian, preventing it from vanishing even at $k_y=0$. A similar behavior with small differences can be found for the case of $B_z=0.6$, as seen in Fig. \ref{Fig6}(d). The non-diverging behavior of $S_x^\pm$ is now attributed to the non-zero $k_x$ acquired by $B_z \neq 0$. The coalescence of $S_x^\pm$ now occurs at non-zero $k_x$, due to the condition presented in Eq. (\ref{Eq14b}). Again, the spin projections degenerate at the same location as the EPs, shown in Figs. \ref{Fig4}(f) and \ref{Fig4}(i), confirming our previous predictions.

When both $B_x$ and $B_z$ are non-zero, as seen in Fig. \ref{Fig6}(e), the denominator of $S_x^\pm$ is affected by $B_y$ and the non-zero $k_x$ acquired by $B_z \neq 0$. Thus, the minimum of real $S_x^\pm$ becomes smaller in this case. The coalescence of $S_x^\pm$ in both real and imaginary parts now coincides with Fig. \ref{Fig5}(b).
 
Finally, we aim to verify the CPL-tuned EPs through the spin projection shown in Fig.~\ref{Fig6}(f). Under the application of right-handed CPL with $\Delta = 0.2$, the spin projections converge at the $(k_x, k_y)$ coordinates defined by Eq.~(\ref{Eq24}), confirming the EP positions. Interestingly, the spin projection pattern under CPL closely resembles the case of $B_z \neq 0$ shown in Fig.~\ref{Fig6}(d), with two key differences. First, the values of $k_x$ and $k_y$ differ due to the distinct coupling mechanism of CPL compared to a static $B_z$ field. Second, the spin orientations are reversed under CPL relative to the $B_z$ case. In other words, in the presence of $B_z$ and within the non-degenerate profile, the positive spin projection is smaller than its negative counterpart, whereas CPL inverts this behavior.
Notably, although $k_x$ in Fig. \ref{Fig6}(f) is smaller than that of Fig. \ref{Fig6}(e), the minimum of Re$[S^\pm]$ in the presence of CPL has weaker signal, that can be attributed to the distinct denominator originated from different coupling of CPL.

\subsection{Opto-magnetic tuning of the EPs}

Up to now, we have considered the proposed junction to be influenced by either optical or magnetic fields individually, in order to avoid structural complexity. We now aim to intuitively analyze the effect of the simultaneous application of both CPL and magnetic fields, referred to as opto-magnetic tuning.

A combination of Eqs. (\ref{Eq1}) and (\ref{Eq22}) with $H_B$ yields the full Hamiltonian of the opto-magnetically tuned FM/UPM junction as 
\begin{equation}
\begin{split}
H^F \approx & t[\textbf{k}^2+\boldsymbol{\alpha}^2+ \frac{2 \Omega \Delta}{\lambda^2}]\sigma_0+ [2t(\textbf{k} \cdot \boldsymbol{\alpha})+2 \Delta+Bz] \sigma_z 
\\&+[\lambda k_y-\frac{2 \alpha_y \hbar^2 \Delta}{m \lambda}+B_x]\sigma_x\\
&-[\lambda k_x-\frac{2 \alpha_x \hbar^2 \Delta}{m \lambda}-B_y]\sigma_y-i \Gamma \sigma_0 - i \gamma \sigma_z.
\end{split}
\label{H with CPL and B}
\end{equation}

The most general conditions for the occurrence of the EPs under the simultaneous influence of CPL and a magnetic field can be written as

\begin{subequations}
\begin{equation}
\begin{split}
&(B_x+\lambda k_y-\frac{2 \alpha_y \hbar^2 \Delta}{m \lambda})^2+(B_y-\lambda k_x+\frac{2 \alpha_x \hbar^2 \Delta}{m \lambda})^2\\
&+(\frac{\hbar^2}{m}(\textbf{k} \cdot \boldsymbol{\alpha})+2 \Delta+B_z)^2= \gamma^2,
\end{split}
\label{Eq28a}
\end{equation}
\begin{equation}
\gamma(2 t \alpha_x k_x+2 t \alpha_y k_y+ B_z+2 \Delta)=0.
\label{Eq28b}
\end{equation}
\end{subequations}
Evidently, the combination of CPL and $\boldsymbol{B}$ affects both conditions. For simplicity, we discuss the effects of $B_x$, $B_y$, and $B_z$, individually. It is clear that combining the components of the magnetic field (e.g., in a planar configuration) results in a merging of their individual effects, as seen in Fig. \ref{Fig5}.

\textbf{Case I}: When $B_x \neq 0$ is applied alongside the CPL, Eq. (\ref{Eq28b}) reduces to (\ref{Eq23b}), but the first term of Eq. (\ref{Eq28a}) still be re-normalized by $B_x$. This re-normalization affects the sign of $\Delta$ term, and hence, changes the ($k_x,k_y$) at which the EPs can be found. More precisely, the EPs now can be found at

\begin{subequations}
\begin{equation}
k_x =\frac{\alpha_x^2 \lambda^3 (\alpha_y B_x - \Delta \lambda) \pm \alpha_y \sqrt{\Lambda }}{\alpha_x \boldsymbol{\alpha}^2 \lambda^4}
\label{Eq29a}
\end{equation}
\begin{equation}
k_y = -\frac{\alpha_x^2 B_x \lambda^3 + \alpha_y \Delta \lambda^4 \pm \sqrt{\Lambda}}{\boldsymbol{\alpha}^2\lambda^4}.
\label{Eq29b}
\end{equation}
\end{subequations}

with

\begin{equation}
\begin{split}
\Lambda=&\alpha_x^2 \lambda^4 [-16 \boldsymbol{\alpha}^2 \Delta^2 + 8 \alpha_y \boldsymbol{\alpha}^2 B_x \Delta \lambda +\\
& (\alpha_x^2 (\gamma^2 - 8 \Delta^2) + \alpha_y^2 (-B_x^2 + \gamma^2 - 8 \Delta^2)) \lambda^2+\\
& 2 \alpha_y B_x \Delta \lambda^3 - \Delta^2 \lambda^4 ].
\end{split}
\label{Eq30}
\end{equation}

At $\alpha_y = 0$, this opto-magnetic modulation results in a combination of the features observed in Figs. \ref{Fig4}(a) and \ref{Fig6}(a). Specifically, applying $B_x$ in the ($E$,$k_y$) diagram tilts and shifts the energy curves, while the CPL controls their separation and modifies $k_x$. Consequently, the position, number, and spacing of the EPs can be tuned via these two parameters. However, in this case, $B_x$ does not influence the critical value of $\Delta$ at which the EPs vanish.
This behavior is similar to what is observed in Fig. \ref{Fig5}(b), when both $B_x$ and $B_z$ are applied, which can be attributed to the analogous roles played by $B_z$ and $\Delta$. Nevertheless, the coupling of $\Delta$ with RSOC leads to distinct ($k_x,k_y$) values at which EPs occur. Still, the angle of EP line could be controlled by appropriate choice of UPM magnet vector strength.

If we consider the UPM region to be $p_y$ wave, i.e., $\alpha_x =0$ and $\alpha_y \neq 0$ is applied, in the ($E-k_y$) plane, the $B_x$ can control the distance of EPs, such that increasing $B_x$ (in the presence of CPL) enhances the distance between two EPs. Thus, now $B_x$ can affect the critical value of $\Delta$ and interestingly enhance it. For instance, the EPs in ($E,k_x$) diagram at $\alpha_y=0.6$ and $B_x=\alpha_x=0$ vanish by applying $\Delta=0.25$, however, increasing $B_x$ up to 0.6, enhances the maximum value of $\Delta$ up to 0.39.

\textbf{Case II}: Considering $B_y \neq 0$ in the presence of CPL, generally changes the EP location to 

\begin{subequations}
\begin{equation}
k_x=-\frac{-\alpha_y^2 B_y \lambda^3 +\alpha_x \Delta \lambda^4 \pm  \sqrt{\Lambda'}}{\boldsymbol{\alpha}^2\lambda^4},
\label{Eq31a}
\end{equation}
\begin{equation}
k_y=\frac{-\alpha_y^2 \lambda^3(\alpha_x B_y+\Delta \lambda) \pm \alpha_x \sqrt{\Lambda'}}{\alpha_y \boldsymbol{\alpha}^2 \lambda^4},
\label{Eq31b}
\end{equation}
\end{subequations}
where $\Lambda'$ can be obtained by replacing $\pm \alpha_x \rightarrow \mp\alpha_y $ and $B_x \rightarrow B_y$ in $\Lambda$ (see Eq. (\ref{Eq30})).

Setting $\alpha_y = 0$ in Eqs. (\ref{Eq28a}) and (\ref{Eq28b}), results in a combination of the features observed in Figs. \ref{Fig4}(c) and \ref{Fig6}(a). More precisely, applying $B_y$ in the ($E$,$k_y$) diagram brings the EPs closer together without changing $k_x$, while CPL plays a similar role but also modifies $k_x$, providing an alternative means to control the position and number of EPs.
Since both $B_y$ and $\Delta$ influence the separation of Eps in this case, the critical value of each field is affected by the other.  Analytically, the critical value of CPL in the presence of $B_y$ for the $p_x$-wave UPM can be written as

\begin{equation}
\Delta=\sqrt{\frac{\alpha_x^2 \gamma^2 \lambda^2}{(4 \alpha_x^2+\lambda^2)^2}}-\frac{\alpha_x B_y \lambda}{4 \alpha_x^2+\lambda^2}.
\end{equation}
For example, in the presence of $B_y = 0.3$ at $\alpha_y = 0$, $\alpha_x = 0.6$, the EPs annihilate for $\Delta > 0.172$, which is lower than the case of $B_y = 0$.

Considering $\alpha_x =0$ and $\alpha_y \neq 0$, the effect of $B_y$ on the ($E,k_x$) diagram is similar to what can be seen for $B_x$-perturbed system in $(E,k_y$) diagram, i.e., the previous case. However, now the EPs are shifted to the right side, instead of left direction. The critical value of $\Delta$ remains unchanged in this scenario.

\textbf{Case III}: When both CPL and $B_z$ fields are applied, condition (\ref{Eq28b}) should be used. This makes $k_x$ dependent on $B_z$ and $\Delta$ even at $\alpha_y = 0$. In the most general case ($\alpha_x \neq \alpha_y \neq 0$), the EP positions can be determined as

\begin{subequations}
\begin{equation}
k_x = \frac{ -\alpha_x (B_z + 2 \Delta) \lambda^4 \pm \sqrt{\zeta} }{ 2 \boldsymbol{\alpha}^2 \lambda^4 },
\label{Eq34a}
\end{equation}
\begin{equation}
k_y = -\frac{
\alpha_y^2 (B_z + 2 \Delta) \lambda^4 \pm \alpha_x \sqrt{\zeta}}{2 \alpha_y \boldsymbol{\alpha}^2 \lambda^4},
\label{Eq34b}
\end{equation}
\end{subequations}

where

\begin{equation}
\begin{split}
\zeta = & \alpha_y^2 \lambda^4 [ 
-64 \boldsymbol{\alpha}^2 \Delta^2 
+ 4 \boldsymbol{\alpha}^2(\gamma^2 - 4 \Delta (B_z + 2 \Delta)) \lambda^2 \\
&- (B_z + 2 \Delta)^2 \lambda^4 
].
\end{split}
\label{Eq35}
\end{equation}


- 
 Since both $B_z$ and CPL affect the separation of EPs in the ($E$-$k_y$) plane, their critical values (at which the EPs vanish) are interrelated. Analytically, the critical value of CPL in a $p_x$-wave UPM influenced by $B_z$ can be derived as

\begin{equation}
\Delta=\sqrt{\frac{\alpha_x^2 \gamma^2 \lambda^2}{(4 \alpha_x^2+\lambda^2)^2}}-\frac{B_z \lambda^2}{2(4 \alpha_x^2+\lambda^2)}.
\end{equation}
As a numerical example, at $\alpha_x = 0.6$, applying $B_z = 0.6$ reduces the critical value of $\Delta$ to 0.12, which is almost half of the case in which $B_z=0$. Thus, manipulating these fields enables control over the number of EPs, as well as their position.
Noting that, the effect of $B_z$ on $(E,k_x)$ diagram at $\alpha_y \neq 0$, is the same as impact of $B_z$ on $(E,k_y)$ diagram at $\alpha_x \neq 0$.

\section{Connect to Experiment}

At the end, we briefly discuss the experimental feasibility of the proposed system. While a direct experimental analog to the specific FM/UPM junction studied here is yet to be reported, there is compelling evidence suggesting that the constituent elements and the predicted phenomena are within experimental reach. The theoretical prediction of $p$-wave magnetism in CeNiAsO \cite{hellenes2023exchange}, corroborated by numerous subsequent theoretical investigations \cite{maeda2024theory,hedayati2025transverse,soori2025crossed,ezawa2024purely,fukaya2025josephson2}, offers a promising material foundation for realizing the UPM component. Very recently, the $p$-wave magnetization is realized experimentally in Gd$_3$Ru$_4$Al$_{12}$ \cite{yamada2025gapping} and in NiI$_2$ \cite{song2025electrical}.

 The fabrication of junctions involving ferromagnetic layers in proximity with non-magnetic semiconductors exhibiting strong spin-orbit coupling, such as InAs \cite{kjaergaard2016quantized} and InSb \cite{gazibegovic2019bottom}, is a well-established practice utilizing standard thin film deposition and lithography techniques. By analogy, the creation of an interface between a ferromagnetic material and a synthesized UPM, leveraging these existing methodologies, represents a logical and feasible step. Furthermore, the study of NH physics in related hybrid systems, such as Josephson junctions based on superconductor-semiconductor heterostructures \cite{nichele2020relating,razmadze2020quantum}, underscores the experimental accessibility of NH phenomena in condensed matter systems.
 
  The crucial coupling between the FM and the UPM, characterized here by $\Gamma_{\uparrow,\downarrow}$, can be effectively controlled through careful manipulation of both the spin-dependent density of states in the FM lead and the tunneling probability across the FM/UPM interface \cite{cayao2023exceptional}. The FM lead provides the necessary spin-dependent density of states, leading to distinct coupling strengths for different spin orientations. Moreover, the overall magnitude of these couplings can be precisely tuned by introducing a non-magnetic potential barrier of finite thickness at the interface. As claimed by Cayao \cite{cayao2023exceptional}, Angle-Resolved Photoemission Spectroscopy (ARPES) presents a viable technique for detecting the predicted EPs in such junctions, owing to the anticipated large spectral features associated with the momenta connecting these degenerate points. Given the established use of ARPES in studying the electronic band structure of similar heterostructures, the experimental observation of EPs in our proposed system appears plausible.

Numerous experimental studies have demonstrated the feasibility of using CPL in the off-resonance regime for manipulating 2D materials \cite{PNAS,feng2019}. When the light is applied in the form of pulses, heating effects can be significantly reduced. As long as the temperature increase due to irradiation remains below the material's critical temperature, the induced magnetic properties are not substantially affected. Consequently, our results remain robust against typical experimental limitations. The CPL energy considered in our analysis reaches 
up to 0.25 $t/a^2$, 
which is comparable to or even lower than values used in previous studies employing similar light sources\cite{APL2021,FFNMoS2,alipourzadeh2022}.

Our theoretical treatment assumes an idealized scenario with a clean interface and the absence of disorder. In practice, non-zero disorder and imperfections may introduce quantitative modifications to the coupling terms and consequently affect the precise locations and properties of the EPs. However, it is generally expected that the qualitative features of the predicted NH physics, including the existence and tunability of EPs via both the intrinsic UPM properties and external fields, should remain robust in the presence of weak disorder, a common occurrence in condensed matter systems. Ongoing advancements in the synthesis and characterization of magnetic materials, particularly those exhibiting the predicted $p$-wave symmetry, pave the way for the experimental realization and comprehensive investigation of the intriguing NH phenomena and EPs predicted in this work for FM/UPM junctions. 
Although direct reports of external controlling the UPM properties are rare, the relevant studies indicates the promising possibility of electrical \cite{chen2024electrical} or strain \cite{chakraborty2024strain} tuning of the AM properties, which could be generalized to the UPMs, but needs more experimental evidences, such as what is done in Ref. \cite{song2025electrical}.
Notably, the RSOC presented in this work also can be tuned by electric field \cite{dash2024role}.

\section{Conclusion}

In summary, we have investigated an NH FM/UPM open junction in the presence of RSOC, influenced by an external magnetic field and CPL as tunable parameters. We demonstrated the emergence of EPs, i.e., simultaneous coalescence of eigenvalues and eigenvectors, whose positions are highly sensitive to the properties of the UPM. Compared to $d$-wave AMs, the EPs in the proposed system exhibit distinct characteristics, such as different numbers and conditions, due to the preserved time-reversal and broken inversion symmetries.
Our results show that both the unidirectional magnetic field (with tunable strength and orientation) and CPL modify the EPs in a momentum-dependent manner, leading to shifts, tilting, merging, or annihilation. While both perturbations reshape the EP structure, their underlying mechanisms differ: CPL induces a global Floquet renormalization, offering dynamic control via light, whereas the unidirectional magnetic field selectively modifies orientation-dependent terms without the same degree of tunability.
When these external fields are applied in a planar configuration, their combined effect closely resembles the superposition of their individual contributions. These behaviors were further confirmed through spin projection textures and eigenvectors overlap calculations. Overall, our findings deepen the understanding of NH physics in UPM-based junctions and highlight the potential of such junctions as versatile platforms for the realization and control of the properties of next-generation spintronic devices.

\textbf{\textit{Note added}}: During the finalization of this work, we noted a related preprint [53] on Néel-vector-controlled exceptional contours; however, our study reveals distinct physics through the introduction of CPL and magnetic field modulations to control the EPs.
\vspace{-12mm}
\begin{widetext}
\appendix
\section{Derivation of the CPL-perturbed effective Hamiltonian}
\label{AppA}
Let us to start from Hamiltonian presented in Eq. (\ref{Eq2}). Considering the potential vector $\textbf{A}(t)=A(\sin (\Omega t),\cos(\Omega t))$ for the right-handed CPL, substituting $\textbf{k}$ with $\textbf{k}-e\textbf{A}$ extends the Hamiltonian as

\begin{equation}
\begin{split}
H_p^{(\mathbf{k} - e\mathbf{A})} &= \frac{\hbar^2}{2m} \left[ \left(k_x - eA \sin(\Omega t)\right)^2 + \left(k_y - eA \cos(\Omega t)\right)^2 + \alpha_x^2 + \alpha_y^2 \right] \sigma_0 \\
&\quad + \frac{\hbar^2}{m} \left[ \left(k_x - eA \sin(\Omega t)\right)\alpha_x + \left(k_y - eA \cos(\Omega t)\right)\alpha_y \right] \sigma_z \\
& \quad + \lambda \left[ \left(k_y - eA \cos(\Omega t)\right)\sigma_x - \left(k_x - eA \sin(\Omega t)\right)\sigma_y \right],
\end{split}
\end{equation}
in which we have also replaced $t$ with $\hbar^2/2m$ to avoid the misleading with the time. Now we can separate the time-dependent and time-independent terms as $H_p^{(\mathbf{k} - e\mathbf{A})}=H_{p,0}^{(\mathbf{k} - e\mathbf{A})}+V(t)$ with the simplified form of

\begin{subequations}
\begin{equation}
H_{p,0}^{(\mathbf{k} - e\mathbf{A})} = \frac{\hbar^2}{2m} \left[ k_x^2 + k_y^2 + \alpha_x^2 + \alpha_y^2 +e^2 A^2 \right] \sigma_0 
+ \frac{\hbar^2}{m} \left[ k_x \alpha_x + k_y \alpha_y \right] \sigma_z 
+ \lambda \left( k_y \sigma_x - k_x \sigma_y \right),
\label{A2a}
\end{equation}
\begin{equation}
\begin{split}
V(t) =& \frac{\hbar^2}{2m} \left[- 2eA k_x \sin(\Omega t) - 2eA k_y \cos(\Omega t) \right] \sigma_0 
\quad + \frac{\hbar^2}{m} \left[ -\alpha_x eA \sin(\Omega t) - \alpha_y eA \cos(\Omega t) \right] \sigma_z \\
&+ \lambda \left[ eA \sin(\Omega t) \sigma_y - eA \cos(\Omega t) \sigma_x \right].
\end{split}
\end{equation}
\end{subequations}

Now, using the van Vleck expansion, the $H_0^F=1/T \int_{0}^{T} H (\textbf{k}-e\textbf{A}) dt$  term becomes equal to $H_{p,0}^{(\mathbf{k} - e\mathbf{A})}$ in \ref{A2a}.


The $H_{\pm1}^F=1/T \int_{0}^{T} H (\textbf{k}-e\textbf{A}) e^{\pm i \Omega t} dt$ term now can be written as

\begin{equation}
H_{(+1)}^F(t) =\ 
\ \frac{\hbar^2}{2m} \, eA \left( -i k_x - k_y \right) \sigma_0 
- \frac{\hbar^2}{m} \cdot \frac{eA}{2} \left( \alpha_y + i \alpha_x \right) \sigma_z 
+ \frac{\lambda eA}{2} \left( i \sigma_y - \sigma_x \right).
\end{equation}

Using $H_{+1}=[H_{-1}]^\dagger$ and the commutation rules of Pauli matrices, the final Floquet Hamiltonian $H^F(k) \approx H_0^F+[H_{-1}^F,H_{+1}^F]/\Omega$ can be derived as we 


\begin{equation}
\begin{split}
H_{\text{eff}} =&\ 
\ \frac{\hbar^2}{2m} \left( k_x^2 + k_y^2 + \alpha_x^2 + \alpha_y^2 + e^2 A^2 \right) \sigma_0 
+ \left[ \frac{\hbar^2}{m} \left( k_x \alpha_x + k_y \alpha_y \right) + \frac{A^2 e^2 \lambda^2}{\Omega} \right] \sigma_z \\
&+ \lambda \left( k_y - \frac{A^2 e^2 \hbar^2}{m \Omega} \alpha_y \right) \sigma_x 
- \lambda \left( k_x - \frac{A^2 e^2 \hbar^2}{m \Omega} \alpha_x \right) \sigma_y.
\end{split}
\end{equation}

This gives Eq. (\ref{Eq22}) by substituting $\Delta=(e A \lambda)^2/2 \Omega$. 
\end{widetext}

\nocite{apsrev41Control}
\bibliography{ref7}
\end{document}